# Examination of the tradeoff between intrinsic and extrinsic properties in the optimization of a modern internal tin Nb₃Sn conductor

C Tarantini[1], P J Lee[1], N Craig[1], A Ghosh[2] and D C Larbalestier[1]

[1]National High Magnetic Field Laboratory, Florida State University, Tallahassee, Florida 32310, USA
[2] Magnet Division, Brookhaven National Laboratory, Upton, NY11973 USA

Email: tarantini@asc.magnet.fsu.edu

In modern Nb₃Sn wires there is a fundamental compromise to be made between optimizing the intrinsic properties associated with the superfluid density in the A15 phase (e.g. $T_c$, $H_c$, $H_{c2}$, all of which are composition dependent), maximizing the quantity of A15 that can be formed from a given mixture of Nb, Sn and Cu, minimizing the A15 composition gradients within each sub-element, while at the same time generating a high vortex pinning critical current density, $J_c$, by maximizing the grain boundary density with the additional constraint of maintaining the RRR of the Cu stabilizer above 100. Here we study these factors in a Ta-alloyed Restacked-Rod-Process (RRP®) wire with ~70 μm diameter sub-elements. Consistent with many earlier studies, maximum non-Cu $J_c$(12 T, 4.2 K) requires preventing A15 grain growth, rather than by optimizing the superfluid density. In wires optimized for 12 T, 4.2 K performance, about 60% of the non-Cu cross-section is A15, 35% residual Cu and Sn core, and only 5% a residual Nb-7.5wt.%Ta diffusion barrier. The specific heat and chemical analyses show that in this 60% A15 fraction there is a wide range of $T_c$ and chemical composition that does diminish for higher heat treatment temperatures, which, however, are impractical because of the strong RRR degradation that occurs when only about 2% of the A15 reaction front breaches the diffusion barrier. As this kind of Nb₃Sn conductor design is being developed for sub-elements about half the present size, it is clear that better barriers are essential to allowing higher temperature reactions with better intrinsic A15 properties. We present here multiple and detailed intrinsic and extrinsic evaluations because we believe that only such broad and quantitative descriptions are capable of accurately tracking the limitations of individual conductor designs where optimization will always be a compromise between inherently conflicting goals.

## 1. Introduction

It has been qualitatively understood for many years that the large compositional variation possible within the A15 Nb₃Sn phase makes optimization of any particular Nb₃Sn conductor an often uncertain compromise between maximizing the amount and quality of the A15 phase in the non-Cu cross-section and minimizing the effective sub-element size, $d_{eff}$, whilst also avoiding any breakdown of the diffusion barrier that would degrade the electrical and thermal conductivity of the stabilizing Cu surrounding the sub-elements. What makes the best compromise uncertain is that the reaction to make the A15 phase from Sn and Nb constituents within the conductor is very dependent on small details of the local Sn, Nb and Cu (i.e. the non-Cu fraction) environment and the reaction time and temperature. Thus progress towards a full optimization of Nb₃Sn conductor design has been slow and empirical, typically proceeding by optimizing one parameter at a time, then followed by finding an acceptable balance of properties in secondary optimizations that maintain as much of the primary optimization as possible. Over the last ~15 years the primary accelerator-physics goal has been to raise the non-Cu $J_c$(12 T, 4.2 K) above 3000 A/mm² so that efficient quadrupole and dipole magnets can be designed [1, 2, 3]. Since field quality (largely determined by the effective filament diameter, $d_{eff}$, after reaction) and magnet protection (largely determined by the residual resistivity ratio, RRR, of the Cu stabilizer) are both of great importance, small $d_{eff}$ and high RRR enabled by effective diffusion barriers now provide the two principal constraints on the architecture and the reaction window that can be applied to the conductor. For accelerator saddle magnets, it is clear that bronze route conductors cannot deliver a high enough non-Cu $J_c$, even though they can provide a good balance of non-Cu $J_c$(12 T, 4.2 K) ~1000-1200 A/mm² and $d_{eff}$ < 5 μm suitable for low hysteretic loss ITER conductors. For accelerator magnets, internal tin (IT) and powder-in-tube (PIT) conductors presently represent the state of the art. Broadly speaking it appears that the best IT conductors of the Rod-Restack-Process (RRP®) design can easily provide non-Cu $J_c$(12 T, 4.2 K) > 3000 A/mm$^2$ but struggle to produce long-piece length conductors with $d_{eff}$ < 35-40 µm [4], while PIT conductors struggle to produce non-Cu $J_c$(12 T, 4.2 K) exceeding 2500 A/mm$^2$, although they do offer a slightly smaller sub-element size [5]. The present paper seeks to understand the way that a balance of intrinsic ($T_c$, $H_{c2}$, Sn content, …) and extrinsic (partition of real estate between Sn, Cu, Nb(Ta) and diffusion barrier, A15 grain boundary density, $J_c$, …) factors controls the properties of a modern high performance RRP® conductor designed for demanding accelerator use, while a subsequent paper will explore some aspects of the same issue for recent PIT conductors.

The further development of high performance Nb$_3$Sn should be enabled by a more detailed understanding of the A15 composition, grain morphology, wire structure and strain that until now has been evaluated in a mostly *ad hoc,* empirical way. In fact the optimization of modern Nb$_3$Sn conductors in terms of their Cu:Nb:Sn ratio, effectiveness of doping (Ti or Ta), type and thickness of the diffusion barrier (DB) necessary to maintain a high RRR has been conducted principally by measuring the in-field transport $J_c$, magnetization, and RRR performance on multiple billets with continuously evolving conductor configurations [2, 6, 7, 8, 9, 10, 11, 12]. An important factor that determines the superconducting properties of Nb$_3$Sn conductors is the distribution of properties developed during the A15 phase reaction. In particular it is well known that properties like critical temperature, $T_c$, and upper critical field, $H_{c2}$, are strongly suppressed by off-stoichiometry compositions [6,13,14] degrading also the critical current density, $J_c$. The way that this averaging affects the whole layer properties has been modeled by Cooley *et al.* [15] and compared to the properties of PIT wires by Godeke *et al.* [11].

In this paper we investigate both the local and the average properties of a state-of-the-art RRP® Nb$_3$Sn conductor from the viewpoint that understanding the distribution of properties is essential to improving their optimization. A vital tool is the specific heat ($C_p$) which can provide the real $T_c$ distribution on a scale of the superconducting coherence length ξ over the whole sample, unlike magnetic measurements, which suffer from screening of lower $T_c$ internal regions by external higher $T_c$ regions and do not properly show the transition broadening [16, 17,18,19]. Here we provide an extended characterization of both intrinsic and extrinsic properties by specific heat, magnetization and detailed microstructure on both the macro- and the micro-scale of the same Nb$_3$Sn wire after different heat treatments (HT) that yield conductor performance that varies from acceptable to degraded [8]. Not unexpectedly, many of the conclusions about the role of A15 grain size and reaction conditions are qualitatively similar to earlier studies cited above. What we believe to be new is the quantitative detail of our study and incorporation of diffusion barrier study as a central focus. Since the state of the art for these RRP® designs is now at sub-elements about half the size of the present conductor, sizing the barrier is crucial for permitting high $J_c$ without unacceptable RRR loss. We believe that the quantitative optimization of such conductors will be greatly helped by applying a suite of characterizations like those presented here to these new, finer filament conductors of the 127, 169 and 217 stacks, where maintaining high $J_c$ and high RRR becomes progressively more difficult. It is our hope that this quantitative baseline characterization will make it easier too to compare recent work on PIT and other variants that vie with RRP® to become the next, best design of advanced Nb$_3$Sn conductor.

## 2. Experimental details

In this paper we study an RRP® wire manufactured by Oxford Instruments, Superconducting Technology (wire RRP-8220) previously characterized by transport in ref. [8]. It has a 0.7 mm diameter and is composed of 54 sub-elements in the 61 stack design shown in figure 1. Each ~70 µm diameter sub-element was assembled with Cu-clad Nb-7.5wt.%Ta (4 at.%Ta) diffusion barrier (DB) and with the same composition of Nb-Ta rods dispersed in a Cu matrix that surrounds a Sn-rich core [20]. The Sn diffusion stages of the heat treatment occurred at 210°C/48 h+400°C/48 h, followed by a high temperature A15 reaction heat treatment at temperatures ranging between 620°C and 750°C for 48-192 h. The samples will here be identified by their final reaction (e.g. 650°C/96h is the wire treated at 650°C for 96 h) and heat treatment will be abbreviated to "HT".

Microscopy was performed in a Zeiss 1540 Crossbeam® Field Emission Scanning Electron Microscope (FESEM) -Backscattered Electron Detector (BSE) using both metallographically polished transverse cross-sections and fracture cross-sections to evaluate wire macro- and micro-structure. Energy Dispersive Spectroscopy (EDS) was performed at 15 kV using standard-less analysis (EDAX TEAM V3-4) with an EDAX

Apollo XP SDD detector. Image analysis was performed primarily using ImageJ software [21]. Grain size distributions were found to be log-normal, thus all measurements reported here use log-normal statistics. Magnetization hysteresis loops were performed in a vibrating sample magnetometer (VSM) in a 14 T superconducting magnet in order to estimate the non-Cu critical current density $J_c$, the pinning force density $F_p$ and the extrapolated irreversibility field $H_{Irr}$ at 4.2 K. The specific heat of the wires was measured at 0 and 16 T in a 16 T PPMS with 25-30 mg samples using a relaxation technique. Each wire was measured without removing the external Cu stabilizer in order to prevent changes in Nb$_3$Sn properties due to release of the precompression applied to the sub-elements by differential thermal contraction of the Cu matrix.

## 3. Detailed Microstructural and Microchemical Description

In this section we present a detailed quantitative description of many aspects of the conductor with a view to understanding how much A15 phase is formed, how much residual core and diffusion barrier remains after reaction, how the residual thickness of the barrier varies and how this correlates to the measured RRR. We also measure the A15 grain size, measure the A15 composition gradient and provide information about the trapped phases within the A15 layer. As noted above we believe that this information will be essential to understanding the present gradual degradation of properties in RRP® conductors as sub-elements are pushed to smaller sizes.

*3.1 Digital macro analysis at the sub-element scale*

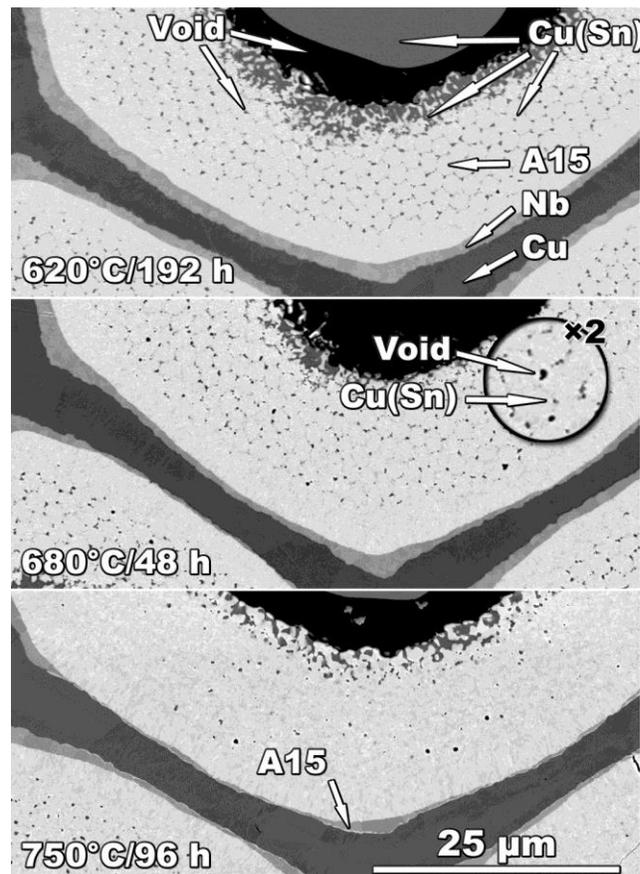

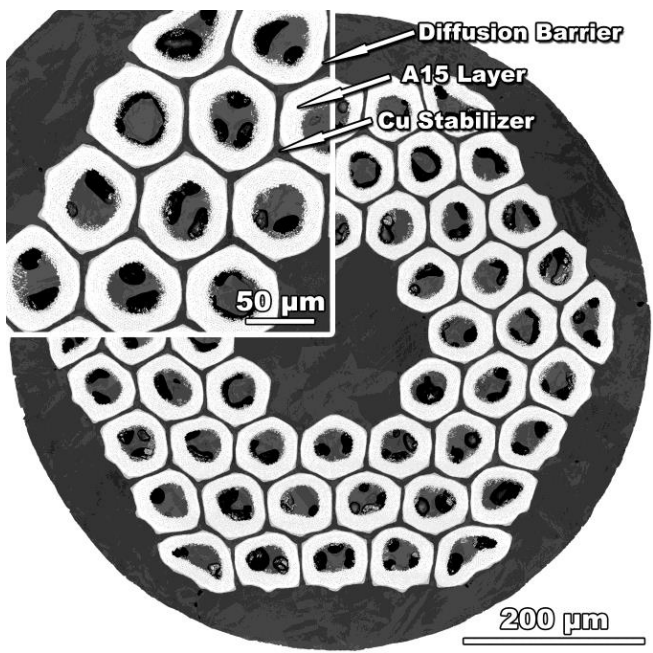

**Figure 1** FESEM-BSE (Field Emission Scanning Electron Microscope-Backscattered Electron) image of a transverse cross-section of the reacted (665°C/50h) RRP® 8220 wire showing the three-ring architecture of the conductor and the tendency for sub-element shape to degrade on going from the inner to the outer ring, where the corners are particularly strongly distorted.

**Figure 2** FESEM-BSE images of transverse cross-sections of three final heat treatments made at progressively higher temperature [620°C/192h (top), 680°C/48h (middle), 750°C/96h (bottom)] which yield decreasing interfilamentary contrast and density of Cu(Sn) islands within the annular filament pack that develops around the Sn core. The unreacted barrier decreases in thickness with increasing A15 reaction and visible A15 reaction appears outside the barrier after the 750°C heat treatment (HT). A non-linear contrast adjustment was made to the images to enhance the interfilamentary contrast and visibility of the A15 layer found outside the barrier.

Figure 1 presents a reacted cross-section of the wire and its 54 non-Cu sub-elements. On the outside of each sub-element is the Nb-Ta diffusion barrier designed to retain the Sn within the sub-element and avoid degradation of the RRR of the stabilizing Cu. After reaction, the originally discrete Cu-clad Nb-Ta filaments form one multiply connected A15 phase annulus ($d_{eff}$ thus approximates the annular diameter of the sub-element) containing a small void density, trapped Nb-Ta and residual Cu that has not counter-diffused into the core. At the center of each sub-element, the original location of the Sn, large voids and a solid solution of Cu and the residual Sn are found (the amount of residual Sn varies with heat treatment and is reported later). The approximate area fractions of these 3 components after reaction are very different, the diffusion barrier being ~5%, the A15 annulus ~60% and the residual core ~35% in the optimally reacted condition. For the purposes of this study the A15 area is measured as the whole connected A15 layer area lying between the diffusion barrier and the Sn core.

A detailed analysis was performed for each wire using a large matrix of FESEM-BSE images (typically montages of 12,000 × 12,000 pixels) which allowed us to analyze the entire cross-section with sufficient resolution to separate voids and interfilamentary Cu(Sn) islands from the A15 layer. In figure 2 we compare FESEM-BSE image details at the same sub-element location in the second sub-element ring for three different HTs of increasing extent. For 620°C/192h there is a large quantity of interfilamentary Cu (Sn) islands and a few voids, as well as a strong interfilamentary contrast. The interfilamentary regions are too thin for a precise chemical analysis by FESEM but their dark contrast indicates lower atomic number than the filament interiors (qualitative EDS mapping will be reported later in figure 11). Increasing the A15 reaction temperature lowers the interfilamentary contrast and increases the void density in the filament pack of the 680°C/48h wire. At the highest reaction temperature (750°C/96h) almost no interfilamentary contrast or Cu(Sn) islands are observed in the A15 annulus and few voids are visible either, implying homogenization of the initially separated filaments and almost complete ejection of the Cu to the (now much depleted) Sn core region [22]. Figure 2 also shows the progressive decrease in thickness of the unreacted Nb-Ta barrier with increasing temperature and the visible A15 reaction that appears on the outside of the barrier after the 750°C heat treatment due to leakage of Sn into the Cu and long range migration through the stabilizing Cu to the reaction site.

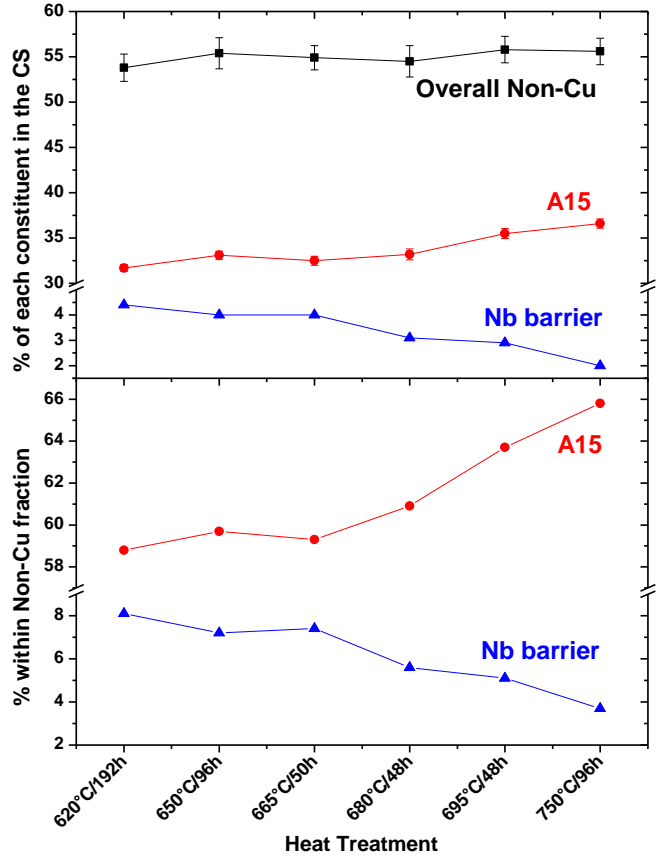

**Figure 3** The heat treatment dependence of (top panel) Non-Cu, A15 and Nb percentages within the wire cross-section and (bottom panel) A15 and Nb percentages within the Non-Cu cross-section of each sub-element.

Table 1 Summary of geometrical measurements for each heat treatment as determined by analytical microscopy (see text for details) and residual resistivity ratio (RRR) measured by transport. Wire diameter (Wire diam.), copper/non copper ratio (Cu/Non-Cu) in the wire, percentage of the wire cross-section (CS) occupied by the sub-elements (Sub.el. % of CS) and its variation (Coeff of. Var.in Sub-el. CS), percentage of the wire cross-section containing A15 (A15,% of CS) and its variation (Coeff. Var. A15), percentage of A15 in the sub-element (A15 of Non-Cu (Sub-el)), percentage of unreacted barrier in the wire and sub-element cross-section (Unreacted barrier, % of wire CS and Unreacted barrier, % of Sub-el.).

| Heat Treatment | Wire diam., μm | Cu/Non-Cu | Sub-el. % of CS | Coeff Var. in Sub-el. CS, % | A15, % of CS | Coeff. Var. A15, % | Unreacted barrier, % of wire CS | A15 of Non-Cu (Sub-el), % | Unreacted barrier, % of Sub-el. | RRR[a] |
|---|---|---|---|---|---|---|---|---|---|---|
| 620°C/192h | 720 | 0.858 | 53.8 | 2.79 | 31.7 | 1.24 | 4.4 | 58.8 | 8.1 | 377 |
| 650°C/96h | 727 | 0.805 | 55.4 | 3.09 | 33.1 | 1.39 | 4.0 | 59.7 | 7.2 | 233 |
| 665°C/50h | 722 | 0.822 | 54.9 | 2.43 | 32.5 | 1.55 | 4.0 | 59.3 | 7.4 | 171 |
| 680°C/48h | 724 | 0.834 | 54.5 | 3.19 | 33.2 | 1.83 | 3.1 | 60.9 | 5.6 | 109 |
| 695°C/48h | 714 | 0.791 | 55.8 | 2.64 | 35.5 | 1.58 | 2.9 | 63.7 | 5.1 | 56 |
| 750°C/96h | 713 | 0.799 | 55.6 | 2.62 | 36.6 | 1.39 | 2.0 | 65.8 | 3.7 | 15 |
| Std. Dev. | 5.4 | 0.025 | 0.8 | | 1.9 | | 0.9 | 2.8 | 1.7 | |
| Coeff. Var.% | 0.7 | 3.1 | 1.4 | | 5.6 | | 26.7 | 4.5 | 26.9 | |

[a] RRR data from ref.8.

Geometrical measurements after the 6 different heat treatments are compared in table 1 and figure 3. As the heat treatment temperature increased so does the A15 phase % in each sub-element (from 58.8% to 65.8%). The residual Nb-Ta barrier area decreased from 8.1% down to 3.7%, but the non-Cu area stayed remarkably constant (notice the difference between gain in A15 and loss of Nb-Ta volumes: the explanation for the volume discrepancy is beyond the scope of this paper but it is discussed for instance in ref. [23]). The A15 fraction within each sub-element is remarkable independent of the sub-element location. For instance, the 665°C/50h wire (figure 1) has an average of 59.3% A15, but the sub-element to sub-element variation is only 0.9 %, despite the strong deformation of the outer corner sub-elements.

*3.2 Diffusion barrier analysis*

The integrity of the Nb-Ta diffusion barrier is essential to maintaining good RRR so we also made detailed measurements of the barrier thickness around each sub-element of each sample, as summarized in figure 4. The Nb(Ta)/A15 and Nb(Ta)/Cu interfaces were identified by intensity thresholding of the atomic-number sensitive backscattered electron images; however, additional manual correction was required in some cases where there were thin external A15 layers resulting from Sn-leakage. The barrier thickness was determined from the digital FESEM images by measuring the distance from every pixel at the Nb(Ta)/A15 interface to the closest pixel at the Nb/Cu interface, using the same technique used in previous position sensitive measurements [24] (a pixel corresponds to 61 nm at the magnifications we use). This is effectively the minimum outwards diffusion distance across the barrier, and because of the geometry of the sub-elements, is slightly smaller than the distance measured from the outside of the barrier to the Nb(Ta)/A15 interface. The data obtained by this technique (and their averages) are used in figure 4c. We found that the barrier reaction extent is quite position sensitive, thinner barriers being found in the more distorted outer ring sub-elements. This is especially so when we compare the length of barrier that has completely reacted through (for instance in the 665°C/50h wire, 0.02% for the inner ring vs. 0.41% for the distorted outer corner sub-elements). The direct relation between the amount of fully reacted diffusion barrier and the degradation of the RRR is evident in figure 4(a). The surprising result of our analysis is that degradation of RRR below 100 occurs when only about 2% of the barrier perimeter has reacted through. The dependence of the barrier reaction on HT is shown in figure 4 (b) which reveals that react-through is limited to only 0.17% up to 665°C/50h but it then rapidly increases to 23% for 750°C/96h. The mean barrier thickness decreases from ~1.2-1.4 μm below 665°C/50h to 0.5-0.6 μm in the 750°C/96h sample (figure 4(c)). Figure 4 presents the challenge for conductor design very clearly: although average thickness is still 0.5-0.6 μm after 750°C/96h, it is the local thinning in just a small fraction of the total barrier perimeter that drives RRR to an unacceptable value of 15.

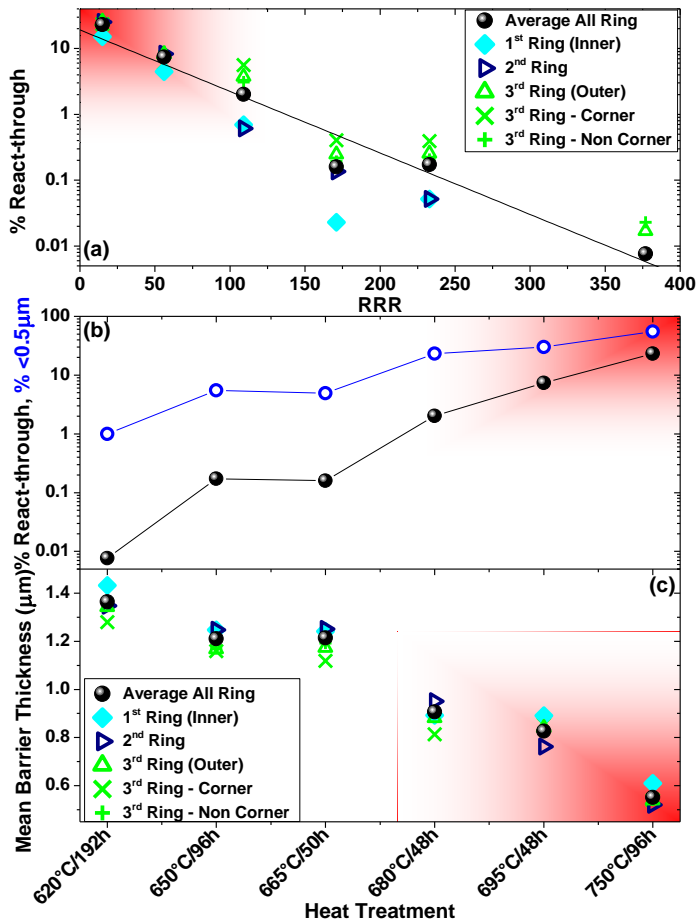

**Figure 4** Analysis of the diffusion barrier integrity: (a) percentage of A15 react-through to the Cu as a function of the RRR value [8]: average thickness, ring by ring, and location within the 3$^{rd}$ ring data are shown; (b) percentage of react-through and percentage of barrier thinner than 0.5 μm as a function of final HT; (c) mean barrier thickness as a function of final HT: average, ring by ring, and location within the outer, 3$^{rd}$ ring data are shown. The shadowed area indicates the wires with significant reduction of barrier thickness and compromised RRR.

Figure 5 shows that there is a strong trend to thicker residual barrier as the distance from the center of the sub-elements increases (we have chosen to show the results just for the 680°C/48h as a mid-range HT example with RRR = 109 but the results for the other HTs are similar). This trend may be due to the outward distortion of the sub-elements that also produces thicker A15 layers on the sides of the sub-elements furthest away from strand center (see for instance figure 1). It is also possible that there is less reaction of the barrier further away from the Sn core, however, confirmation of these hypotheses will require a study of unreacted strand that was not covered in this work.

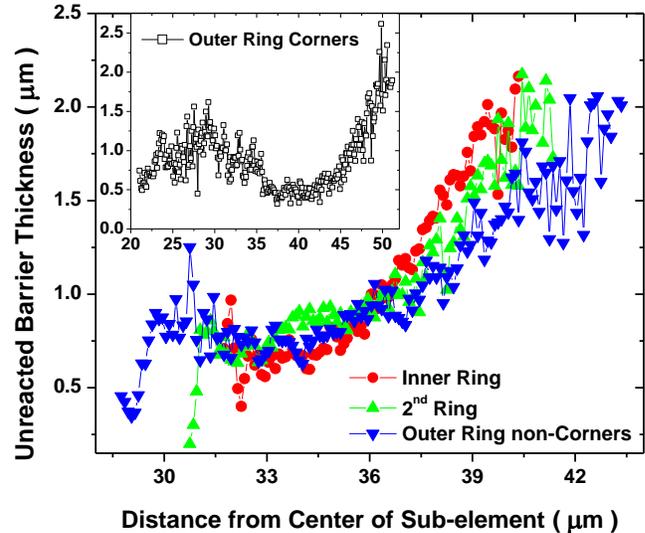

**Figure 5** Unreacted barrier thickness plotted against distance to the outside surface of the barrier from the center of each sub-element for the 680°C/48h heat treated strand. The trends to thicker unreacted barriers with increasing distance from the center of the barrier are similar for all sub-elements except the 6 highly distorted outer ring sub-elements (inset) which show a more complex trend.

*3.3 A15 grain size analysis*

Although the filament pack within each sub-element is almost fully reacted, there are some locations where the filaments themselves do not fully react, as shown in figure 6. The length and duration of the final heat treatment do not change the number or distribution of these partially reacted areas. Although these areas have a negligible impact on the total A15 area, they are of interest with regards to the progress of Sn diffusion. As shown in figure 6, these regions occur in similar regions of the 2$^{nd}$ and 3$^{rd}$ ring of sub-elements suggesting a consistent mechanism by which the Sn diffusion needed for full reaction is blocked by inhomogeneous sub-element deformation during wire fabrication. FESEM-EDS mapping from a partially reacted region of the 620°C/192h sample (bottom right figure 6) is shown in figure 7, confirming the presence of the unreacted filament cores. The EDS maps also indicate a Ta-poor/Sn-rich ring around each of the original filament locations. The Ta-poor ring becomes less pronounced as the distance increases from the Sn core. The Sn-rich regions appear to be thicker around the partially reacted filaments and there are fewer Cu islands. Higher magnification mapping from fully reacted regions will be reported later. Unreacted Nb regions were previously discussed by Pong *et al*. [25] but, differently from our case, they were found near both the core and the diffusion barrier. The reason for this blocking of Sn diffusion is still unknown and it may be similar in the two cases, but a further investigation of intermediate heat treatments, like that performed in ref. [25] would be helpful to clarify this phenomenon.

Figure 8 compares the microstructures revealed by fractography for the 620°C/192h and 680°C/48h wires. In both cases, the initial location of the Nb filaments can be recognized (white circles), although this is more evident at lower reaction extents such as 620°C/192h, where the aspect ratio of the grains and their radial distribution is more pronounced. In both wires a 1.5 µm-thick layer of columnar grains at the far side of the reaction front close to the Nb barrier is visible. Similar images have been taken on all wires in order to estimate the A15 grain size, grain aspect ratio and grain boundary (GB) density. The typical trend of aspect ratio and grain size across the A15 layer is shown in figure 9. It shows that the mean A15 grain diameter $d^*$ (the geometrical diameter of the grains obtained from their cross-sectional areas if assumed to be circular) is quite uniform in the small grain (100-200 nm diameter) region that forms the large majority of the layer. However the grain size is 3-5 times larger close to the Sn-rich core, growing a large-grain A15 layer reminiscent of the inner annulus of much larger A15 grains found in PIT conductors [12]. The grain aspect ratio undergoes a significant increase far from the Sn source near the Nb barrier, changing morphology from equiaxed to columnar. Reflecting the grain growth that occurs with increasing HT temperature, the A15 GB densities progressively decrease from $2.64 \times 10^7$ m$^{-1}$ to $1.4 \times 10^7$ m$^{-1}$ on going from 620°C/192h To 750°C/96h. These values will be used in section 5 to determine the specific GB pinning force. The fractography does not, however, show the small voids or ductile interfilamentary Cu(Sn) islands that are more readily visible in the images of polished samples in figures 2 and 6.

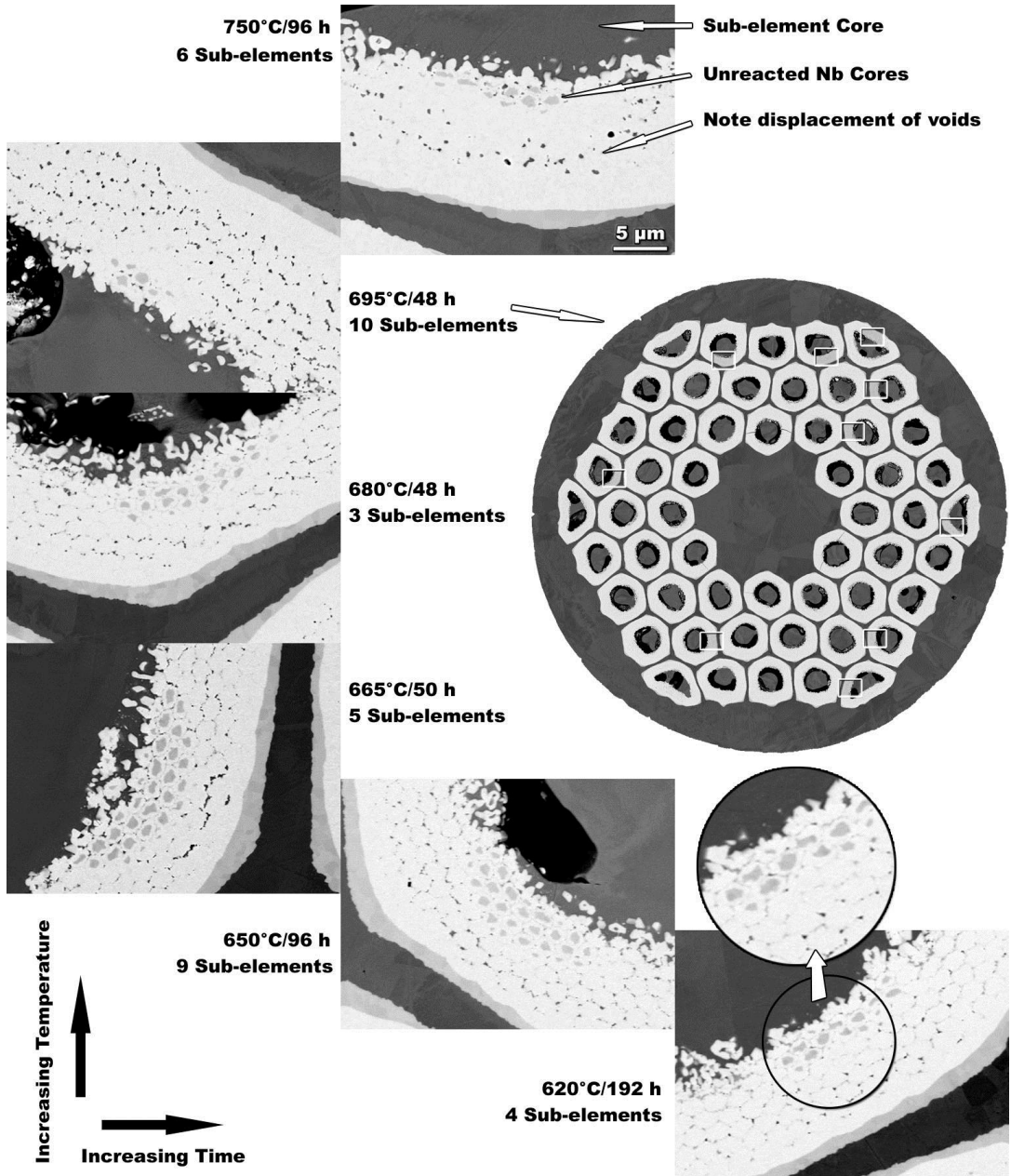

**Figure 6** FESEM-BSE image details for all 6 heat treatments showing examples of unreacted filament centers adjacent to the Sn cores of each sub-element and the number of sub-elements for each heat treatment that have such unreacted filament centers. In each example, particularly for the high temperature heat treatments, the location of the voids has been displaced around the partially unreacted region. We also show the location of the 10 partially reacted regions in the 695°C/48h cross-section at right center.

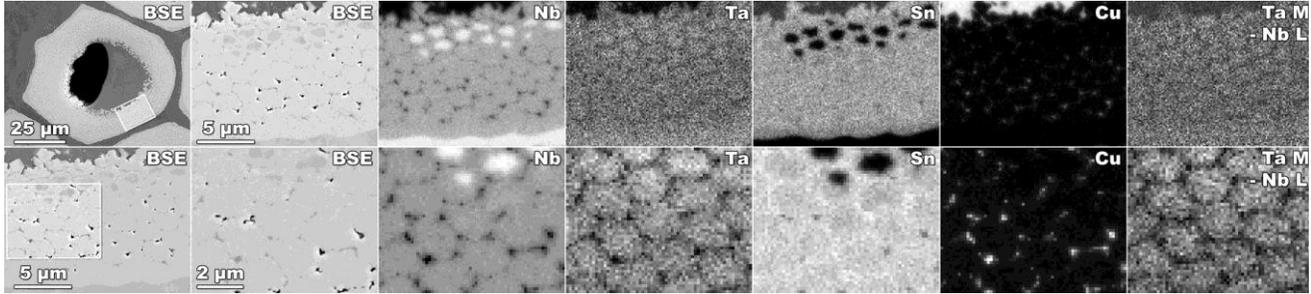

**Figure 7** Qualitative (brighter = higher atomic composition) maps (Nb (L), Ta (M), Sn (L) and Cu (L)) obtained by FESEM-EDS for the partially reacted region of the 620°C strand shown in the bottom right hand corner of Figure 6. The image column on the right subtracts the Nb-L intensity from a normalized Ta intensity (for the upper image the intensity is normalized to the unreacted Nb-Ta barrier and for the lower image the intensity is normalized to the unreacted filament cores). The normalized Ta-Nb maps also show the variation in Ta depletion with distance across the A15 layer.

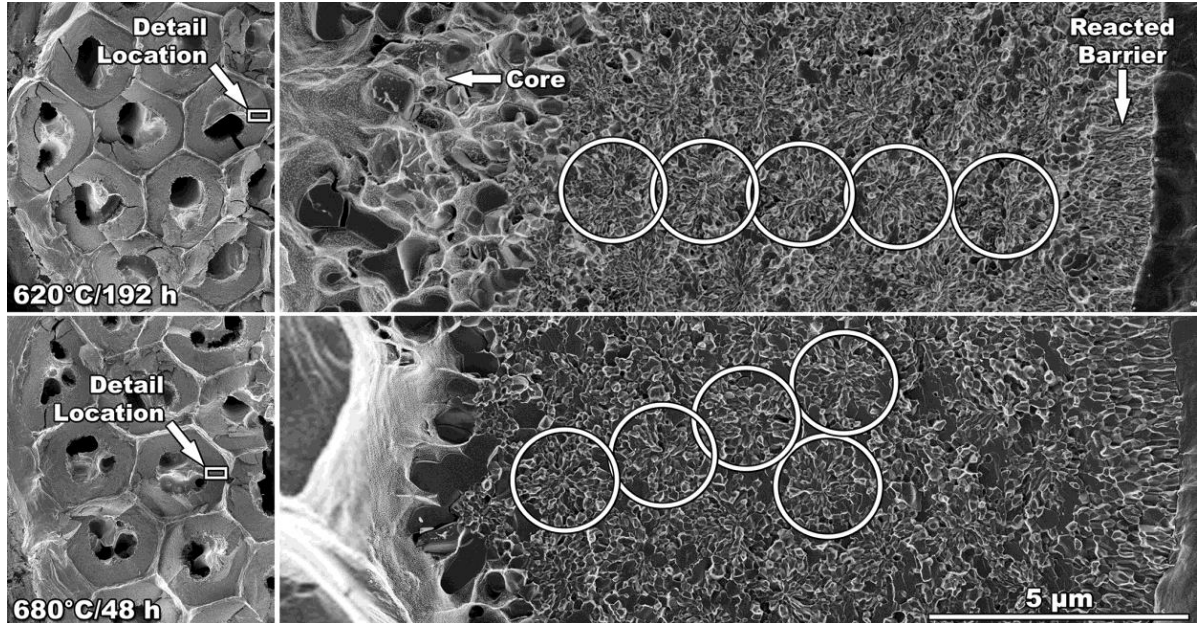

**Figure 8** In-lens secondary electron FESEM images of fractured 620°C/192h and 680°C/48h wires showing the A15 grain microstructure variation across the whole A15 layer (right) and the location of the examined layers (left). White circles identify the approximate centers of some original Nb-Ta filaments. EDS measurements of local Sn and Cu content reported in figure 10 were made both in the approximate centers of filaments from one side to the other of the whole A15 layer and in some cases at the edge of individual filaments so as to observe if there was a radial gradient within each filament. Large disconnected grains are observed on the Sn-rich core side of the A15 layers. A 1.5 µm-thick layer of columnar grains is present next to the unreacted Nb-Ta barrier on the right where the Sn concentration is lowest.

*3.4 Chemical composition analysis*

Microchemical analysis by EDS was performed with the aim of discovering whether there were macroscopic Sn gradients across the whole A15 annular layer and additional local gradients within the regions defined by the original Nb-Ta filaments. For consistency the layer chosen was in the middle ring on the same ring direction side (a compromise between the thicker outer layer and the thinner inner layer) as shown in the schematic illustration under figure 10. Global gradient measurements were made mainly in the center of the original Nb-Ta filaments, where the A15 Sn composition is expected to be locally lowest. Figure 10 shows that the A15 Sn content is relatively flat in the 20-30% of the A15 layer closest to the Sn core for all HTs. At the lowest temperature, 620°C/192h, the Sn composition decreases at larger distances from the Sn core with a strong filament-to-filament variation (21.2-23.4 at.%Sn) that indicates significant reaction inhomogeneity. 680°C/48h has a similar decrease in the Sn composition across the layer but with a much smaller filament-to-filament variation, showing that higher temperature HT favors a more uniform Sn diffusion through the whole filament pack. At even higher temperature (695°C/48h and 750°C/96h), the gradient is even flatter and the A15 more homogeneous. Similar chemical analysis performed on the outer perimeter of filaments in the 695°C/48h case (empty symbols in figure 10) shows about 1.5% higher Sn than in the center of the filaments and values close to stoichiometry (24.5-25 at.%Sn). In the lower panel of figure 10 the presence of Cu in the A15 layer is investigated as well. It is evident that 620°C/192h has much more Cu than the other heat treatments. There is a significant gradient (5.7-2.4 at.%Cu), whereas at the highest HT temperature the gradient disappears and the Cu content ranges between 1.7 and 2.7 at.%. We do not yet know how much of this increased Cu is actually incorporated into the A15 grain or A15 grain boundaries or is from the residual Cu(Sn) islands (of which there are more at lower HT temperatures).

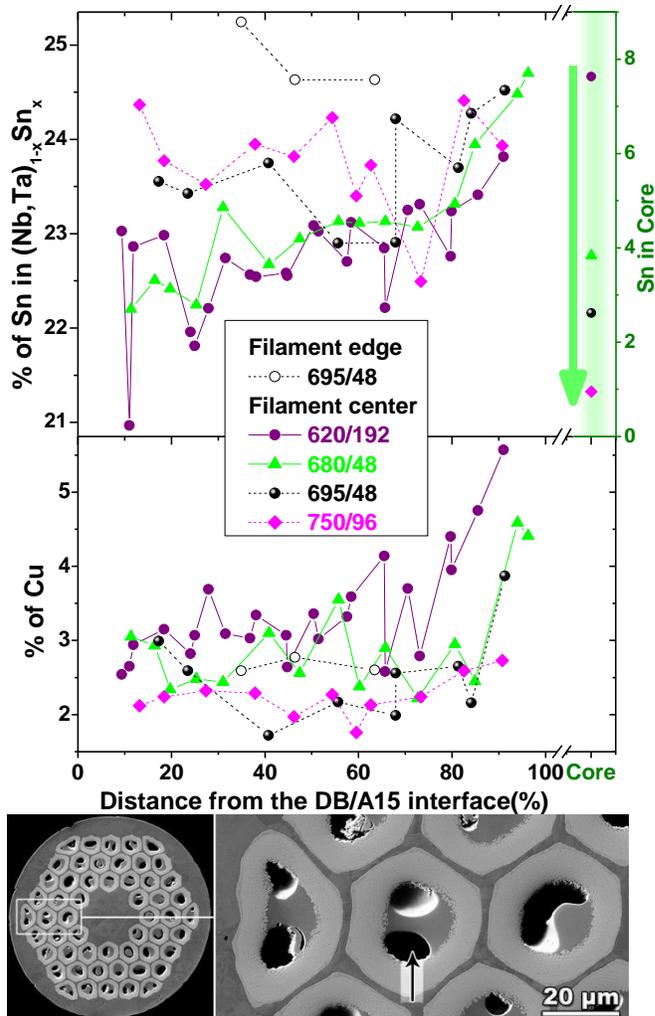

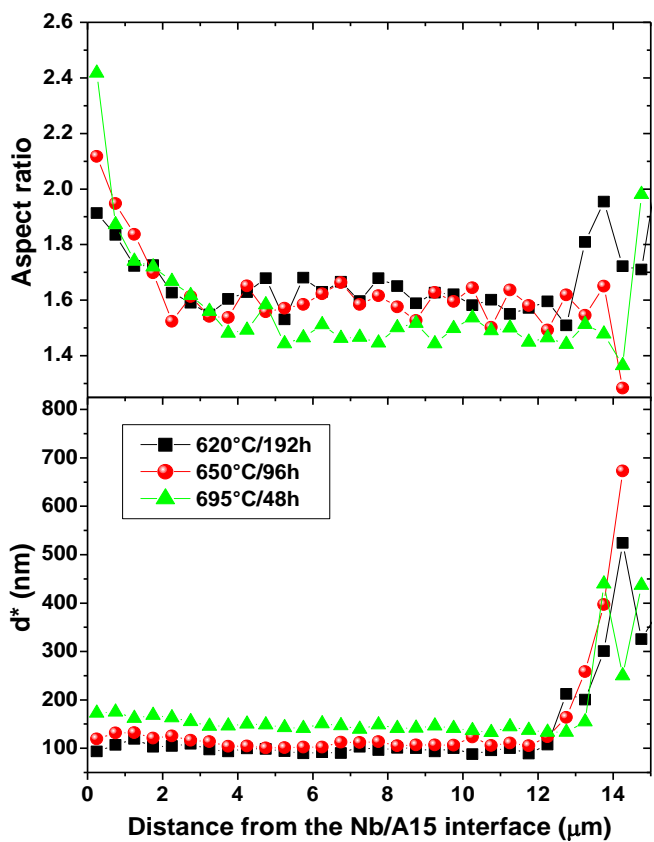

**Figure 9** Aspect ratio (top panel) and A15 mean grain diameter $d^*$ (bottom panel) across the A15 layer starting from the A15/Nb interface. Each data point represents the log-normal mean from a 500 nm wide bin.

**Figure 10** Sn (top chart panel) and Cu (bottom chart panel) atomic compositions across the A15 layer for 4 different HTs determined by EDS (standard-less analysis). The composition has been estimated in the center of the original Nb-Ta filaments (full symbols) or at filament edges in the 695°C/48h wire (empty symbols). The estimated compositional error for the Sn and Cu are approximately ±1.1 and ±0.8 at. %, respectively. These data provide evidence for both filament-scale and whole filament-pack composition gradients. On the right side of the top panel the amount of the residual Sn in the core is also reported, showing a progressive decrease increasing the HT temperature. The EDS location was the same for all strands (2nd ring, ring direction) and is shown the image panel below the chart.

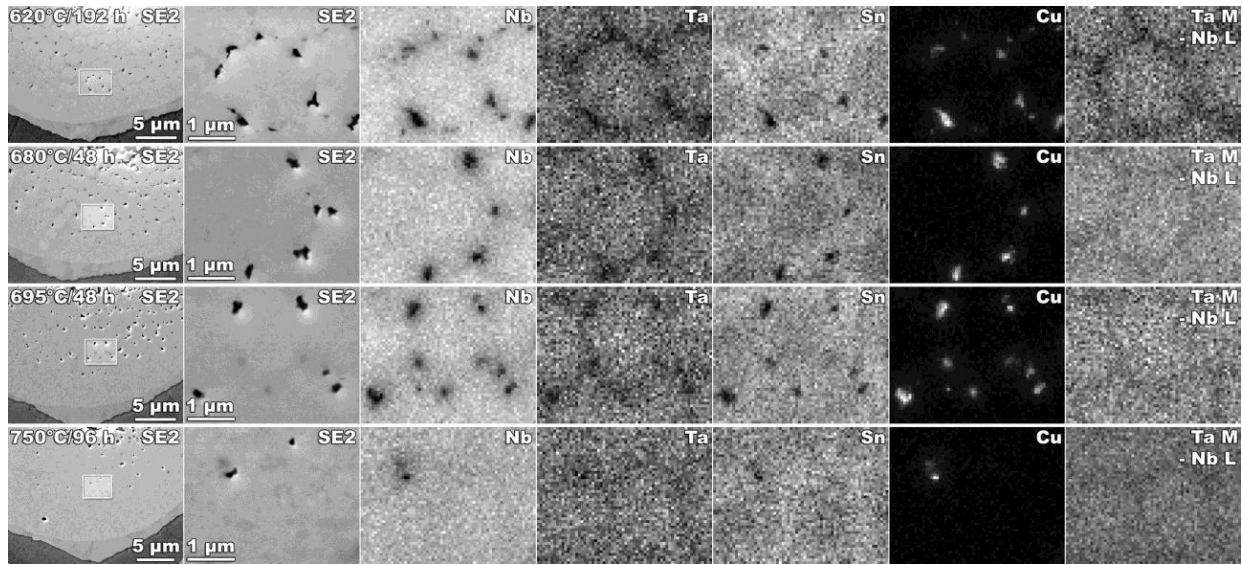

**Figure 11** FESEM-EDS qualitative elemental atomic compositional maps (64x50 data points) showing variations in Nb (L), Ta (M), Sn (L) and Cu (L) at the filament level. The location of the detailed maps is shown in the images on the left column of the panel. The image column on the right subtracts the Nb-L intensity from a normalized Ta intensity (using the same normalization factor used for the bottom right image in figure 7). Each of the original filament locations is surrounded by a high-Sn low-Ta ring creating a cellular looking microchemical distribution. The normalized Ta-Nb maps also show the declining Ta segregation with increasing heat treatment temperature.

The spatial composition variation at the filament level was investigated using FESEM-EDS mapping as is shown in figure 11. The maps indicate compositional variations in the Nb, Ta and Cu that are associated with the original filament locations; there is a low-Ta and a high-Sn ring around each of the original filaments. As the temperature of heat treatment increases the Ta distribution is homogenized, especially when the heat treatment is increased from 695°C to 750°C. However, even at 750° there are distinct Sn-rich and Ta-poor rings. Because the spatial resolution of this technique is only ~1-2 μm it is likely that the real gradients are smeared by the technique and that STEM-EDS will be necessary to discover the true thickness and composition of the layers around the filaments.

In summary, it is clear that the spatial variation of microstructure in this conductor is always measurable and evident, even though it is much reduced by increased reaction. The constraint to maintain RRR > 100 imposes strong requirements on the diffusion barrier for which average residual thickness properties are almost irrelevant since degradation occurs when only 1-2% of the barrier is reacted through to the stabilizing Cu.

## 4. Specific heat characterization and analysis

*4.1 Viewpoint of our specific heat studies*

Having provided a very detailed description of the variability of many important aspects of the macroscopic layout of the conductor and of the amount of A15 phase, its grain size distribution and chemical gradients, we now turn in this and the following section to a description of the superconducting properties to see how they correlate to the previous section.

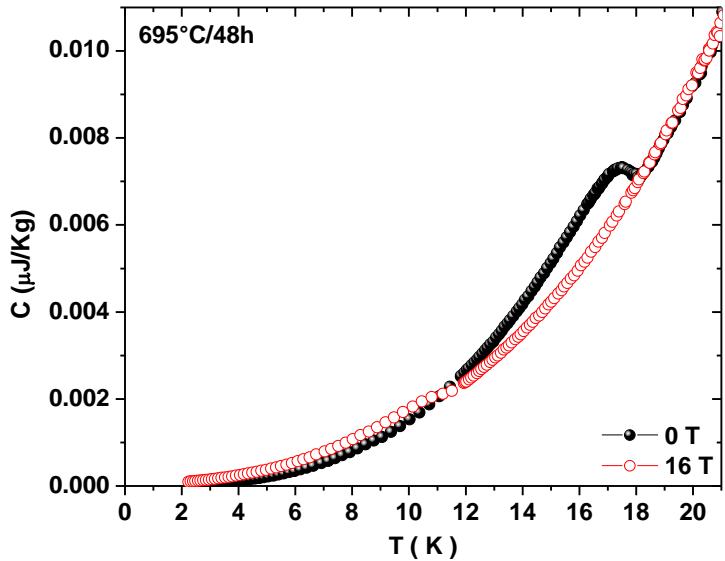

**Figure 12** Specific heat as a function of temperature at 0 and 16 T on an RRP® wire heat treated at 695°C for 48 hours.

Specific heat characterizations are frequently performed on superconducting compounds in order to study properties like $T_c$ and gap amplitude in superconducting state or basic material science properties like structural transitions, electron density and Debye temperature in the normal state. This technique can also provide hints about phase uniformity (from the transition width) in materials with narrow compositional ranges of the superconducting phase and, being a technique sensitive to the whole sample, specific heat can distinguish between a fully and a partially superconducting sample. This technique, however, is rarely employed on wires and tapes in order to investigate the superconducting properties (although it is sometimes employed to characterize the normal state properties useful for magnet design). The case of A15 wires is an exception because of the wide compositional range of the superconducting phase and because all of these compositions are in principle present in a wire. Because of the direct correlation between $T_c$ and the Sn-content, the compositional gradient induces a $T_c$-distribution in the sample that can be evaluated using a model developed by Wang and Senator *et al.* [16, 17] described later. This technique is very effective in determining differences in wires based on different designs, such as the bronze-process, PIT and RRP® [7,17], or different doping methods [26], or to evaluate the presence of distinct A15 layers within the same wire [27]. The question we pose here is whether specific heat measurements are quantitatively useful for studying the systematic changes that occur in a progressive series of reaction heat treatments.

In this study the specific heat measurements were performed both at 0 T in order to determine the $T_c$-distribution, and at 16 T, the highest available field, so as to allow us to better determine the normal state properties (see later discussion and Appendix) and to provide information about the in-field behavior. Figure 12 shows a typical $C(T)$ measurement. At zero-field two features, typical of superconducting transitions, are observed: a first just above ~18 K, corresponding to the onset of the highest-$T_c$ A15 phase fraction, and a second one at about 9 K (less visible in the $C(T)$-plot), related mainly to the superconducting transition of the Nb-Ta diffusion barrier (some residual unreacted Nb-Ta in the filament pack may also contribute: see figure 6). At 16 T the A15 transition is suppressed to about 12 K (red curve in figure 12) and the Nb-Ta transition is completely suppressed. In order to investigate the superconducting properties, the electronic contribution to the total specific heat of the superconducting phases alone ($C_e$) has to be isolated, since this is the only contribution carrying information about the superconducting state [28]. Therefore we need to take into account the presence in the wire of the two superconductors, Nb$_3$Sn and Nb, and normal materials, like Cu and other Nb-Sn-Cu phases. The specific heat data were analyzed using the usual procedure described for instance in ref. [16,17, 29], with a small modification in order to exclude the contribution from Cu and other non-superconducting phases that are not of interest here (see Appendix for more details). Figure 13 reports the obtained C$_e$ contribution normalized to $\gamma T$ (where $\gamma$ is the Sommerfeld constant of the superconducting phases as determined in the Appendix). This plot reveals significant differences in the wires with increasing heat treatment. The onset of the Nb$_3$Sn transition at 0 T ($T_{c,\text{onset}}$ in table 2) lies in the 18.3-18.4 K range for HTs above 665°C but at ~18.1 K for HTs below 650°C. The amplitude of the C$_e$ jump at $T_c$ significantly rises on raising the HT temperature and the position of the maximum moves to higher temperature making the transition sharper. This is particularly evident going from 680°C to 695°C. In contrast, the jump at the Nb transition becomes smaller with increasing HT temperature, indicating greater conversion of the Nb barrier to A15 as it reacts with Sn.

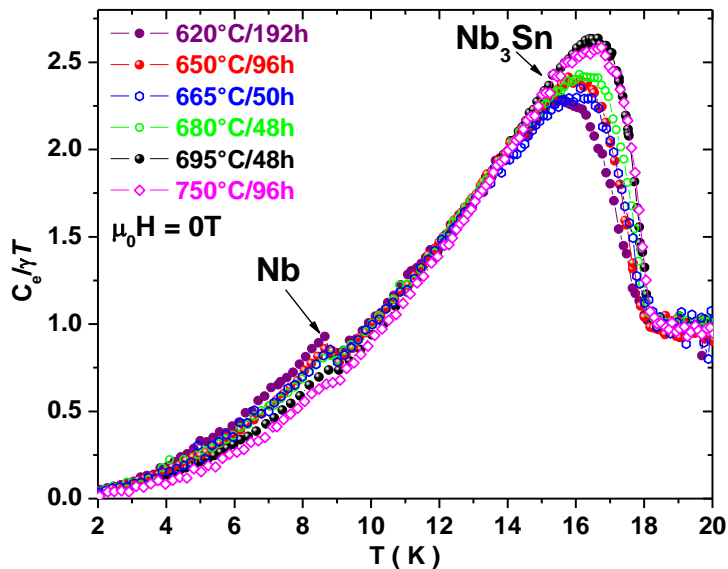

**Figure 13** Normalized electronic specific heat as a function of temperature at zero-field for the same RRP® wire after different heat treatments. The two principal features are the A15 and Nb $T_c$ transitions.

**Table 2** Properties of the wires as determined by specific heat analysis (see details in the text) and RRR measured by transport. Onset ($T_{c,\text{onset}}$), middle point ($T_{c,\text{mid}}$), and transition width ($\Delta T_c$) of the A15 superconducting transition at 0 T obtained from the $F(T)$ curves excluding phases below Nb transition, residual Nb (Res. Nb) as determined by the jump at ~9 K in the $F(T)$ curves, main peak position ($T_{c,\text{peak}}$) and width ($\Delta T_{c,\text{FWHM}}$) obtained from the $f(T_c)$ curves, middle point of the 16 T transition ($T_{c,\text{mid}}(16\,\text{T})$) as determined from the $F(T)$ curves at 16 T, residual phase in superconducting state at 16 T and 2K ($1-F_{2\text{K},16\text{T}}$).

| Heat Treatment | | $T_{c,\text{onset}}$ | $T_{c,\text{mid}}$ | Res. Nb | $\Delta T_c$ | $T_{c,\text{peak}}$ | $\Delta T_{c,\text{FWHM}}$ | $T_{c,\text{mid}}(16\text{T})$ | $1-F_{2\text{K},16\text{T}}$ | RRR[a] |
|---|---|---|---|---|---|---|---|---|---|---|
| Temp °C | Time h | K | K | % | K | K | K | K | % | |
| 620 | 192 | 18.10 | 16.83 | 7.6 | 3.23 | 17.11 | 1.46 | 9.05 | 42 | 377 |
| 650 | 96 | 18.09 | 17.06 | 6.6 | 3.03 | 17.67 | 1.42 | 9.82 | 40 | 233 |
| 665 | 50 | 18.31 | 17.14 | 4.9 | 3.74 | 17.67 | 1.31 | 9.81 | 37 | 171 |
| 680 | 48 | 18.40 | 17.24 | 4.1 | 3.34 | 17.72 | 1.18 | 9.95 | 42 | 109 |
| 695 | 48 | 18.36 | 17.51 | 3.7 | 2.27 | 17.85 | 0.91 | 10.40 | 41 | 56 |
| 750 | 96 | 18.34 | 17.57 | 1.8 | 2.29 | 17.85 | 0.83 | 10.56 | 50 | 15 |

[a] RRR data from ref.8.

*4.2 Deconvolution of the $T_c$ distribution from the measurements*

The C$_e$ data can be deconvoluted to determine the $T_c$-distribution by using the method developed in ref. [16]. This method considers a two-fluid model under the hypothesis that there is one normal state electronic specific heat for all the superconducting phases, independent of $T_c$ ($C_{en}(T,T_c) = C_{en}(T) = \gamma T$), and that the superconducting state electronic specific heat can be described by the relation $C_{es}(T,T_c) = n\gamma T_c(T/T_c)^n$ (according to the Gorter-Casimir model). Moreover the model employs a weighting function $f(T_c)$ that represents the distribution of $T_c$ between 0 and the maximum critical temperature, $T_{c,\text{Max}}$. $f(T_c)$ is related to C$_e$ and to the electronic entropy S$_e$ by the relation [16]:

$$\frac{n S_e(T) - C_e(T)}{(n-1)\gamma T} = \int_0^T f(T_c) dT_c \equiv F(T) \qquad (1)$$

where $F(T)$ is the integral of the $T_c$-distribution from 0 to T. By definition $F(T)$ is constrained by $F(T_{c,\text{Max}})=1$ and it usefully represents the superfluid volume fraction of sample with $T_c \leq T$.

*4.3 Integral of $T_c$-distributions*

The $F(T)$ curves obtained by eq. (1) are displayed in figure 14. Although the $F(T)$ curves appear similar to the magnetization transitions shown in the inset, the specific heat curves actually provide considerably more information because they represent a genuine bulk average of $T_c$ on a scale of the coherence length $\xi$. As expected the two principal features in figure 14 occur at ~9 K and ~18 K and correspond to the Nb and the highest part of the A15 transitions. The width of the Nb transition is only weakly affected by heat treatment (broadening becomes more obvious above 695°C) but the dominant effect is that the Nb transition amplitude progressively decreases to almost zero with increasing reaction temperature. The Nb superfluid volume fraction estimated from the $F(T)$ curves indeed drops from 7.6% at 620°C/192h to 1.8% at 750°C/96h (residual Nb in table 2). In regard to the A15 Nb$_3$Sn transition, HT makes clear differences in both the width and the amplitude. It is particularly notable that much sharper transitions occur for high reaction temperatures. In order to quantify this sharpening, the A15 transition broadening $\Delta T_c$ is defined by the 10-90% criterion on the A15 transition i.e. excluding the Nb transition and the lowest $T_c$ phases]. $\Delta T_c$ decreases from 3.0-3.7 K for HTs below 680°C to ~2.3 K for HT above 695°C (table 2). These $F(T)$ transitions indicate the considerable inhomogeneity of the A15 layer since, even for the highest temperature reaction, over 28% of the A15 phases has a $T_c$ below 17 K whereas for the lowest temperature reaction this amount reaches 57% (excluding Nb and A15 found mostly near the diffusion barrier with $T_c$ below 9 K). The magnetic transition determined for comparison (the 650°C/96h and 695°C/48h curves are shown in the inset as example) reveal quite a different behavior: they show an onset lower than the one determined by specific heat. In fact when the $M(T)$ starts to show a visible diamagnetic signal, the $F(T)$ transition reveals that already ~20% of the A15 phase is in the superconducting state. Moreover, differently from the $F(T)$ case, in the $M(T)$ curves the transition width of the A15 phase and the relative amplitude of the Nb and A15 transitions are similar for these two wires and only a small rise of the A15 transition is observed for the higher temperature reaction. The reason for such different results will be given in the discussion.

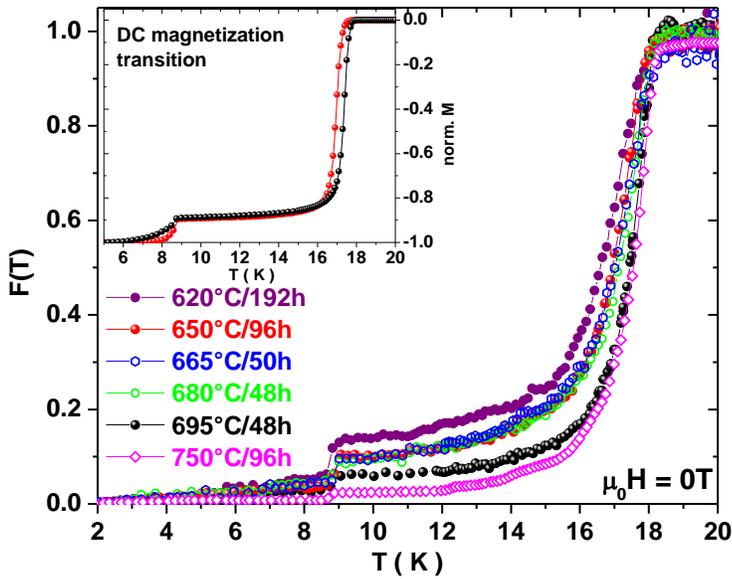

**Figure 14** $F(T)$ is the integral of the $T_c$-distribution at zero field as defined by eq. 1. It represents the volume fraction with $T_c \leq T$. The inset shows the normalized magnetization transition for the 650°C/96h and 695°C/48h heat treatments.

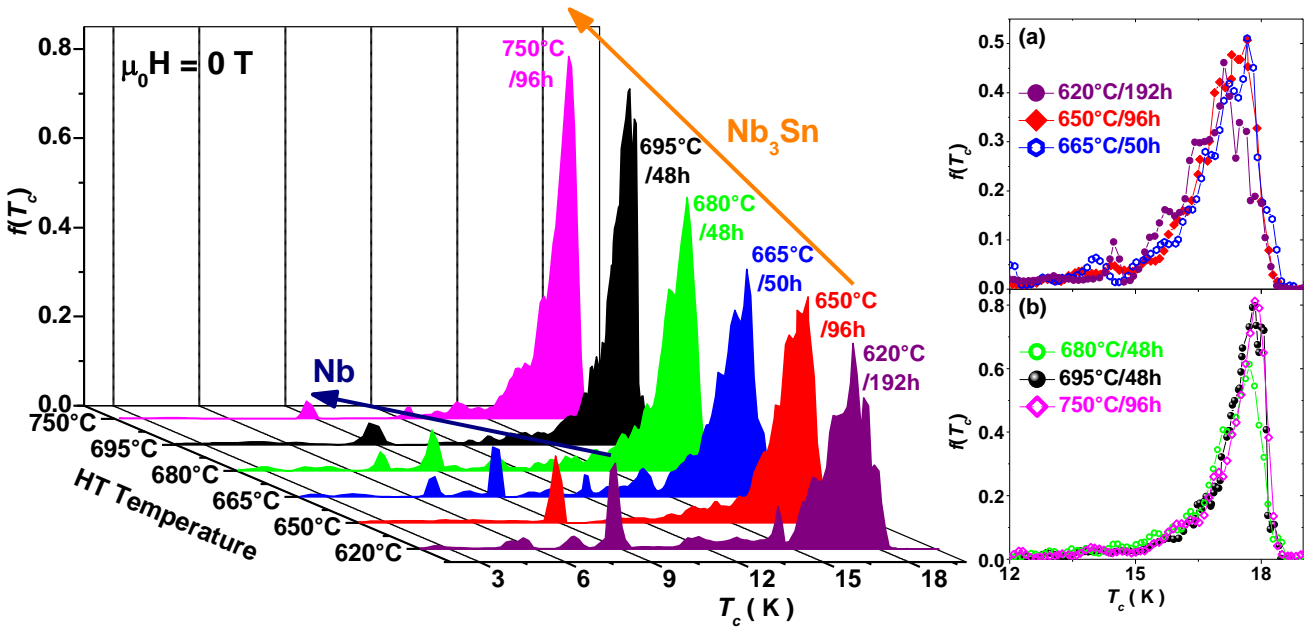

**Figure 15** $T_c$-distributions $f(T_c)$ as a function of heat treatment (HT) temperature and, on the right, a magnified view of the A15 $T_c$ peaks at low (a) and high (b) HT temperature.

## 4.4 $T_c$-distributions

Figure 15 shows the $T_c$-distributions $f(T_c)$ (i.e. the derivative of $F(T)$ shown in figure 14 as follows from eq.1) for the six different HT conditions. In general increasing the HT temperature causes the A15 peak to grow and the $T_c$-distribution to move toward higher $T_c$, whereas the Nb peak becomes smaller. In fact, as summarized in table 2, the $T_c$-distribution peak, $T_{c,peak}$, moves from ~17.1 K to 17.85 K with increasing HT temperature, whereas the peak width decreases from ~1.5 K to ~0.8 K. It is interesting to note that 620°C/192h has the same onset as 650°C/96h, but also that it is clearly more inhomogeneous having multiple peaks above ~14 K in the $T_c$-distribution and the lowest $T_{c,peak}$ (~17.1 K against ~17.7 K of 650°C/96h). Comparing the next two HT conditions (650°C/96h and 665°C/50h) clarifies the effect of HT time in driving more Sn into the A15 layer (figure 15a). In fact the two $f(T_c)$ curves have the same peak intensity and position (~17.7 K) but, despite a higher onset, 665°C/50h has a high temperature shoulder and multiple peaks down to 10 K, suggesting that the longer HT (650°C/96h) drives Sn more effectively into the filament pack and makes an overall more homogeneous A15 layer. Increasing the HT temperature above 680°C causes an increase in the intensity of the highest-temperature A15 peak and makes the secondary peaks in the 10-15 K range almost disappear. Probably this higher HT temperature is removing some spatial variation of reaction associated with small differences in the filament pack arrays and their influence on the Sn diffusion path into each sub-element. The $T_c$-distribution of the two highest-temperature HTs (695°C/48h and 750°C/96h) reveals that the two curves have the same peak position and in the 750°C/96h wire only a weak improvement in peak intensity and width are obtained (figure 15b).

## 4.5 In field specific heat measurements

The initial purpose of performing high-field measurements was to determine the normal state with more accuracy; however they also provide useful details about the in-field $T_c$-distribution. This may be of particular interest for that fraction of the A15 expected to be found near the diffusion barrier with $T_c < 10$ K which is not expected to be superconducting at 16 T. Therefore a similar analysis of the specific heat transitions has been performed at 16 T, seeking correlations with the in-field $J_c$ performance. Figure 16 shows that increasing the HT temperature up to 695°C produces a striking sharpening of $F(T,16T)$ as $T_{c,max}$ is approached. An even more marked change occurs in the 750°C/96h sample. The quantity $1-F(T,16T)$ represents the fraction of the sample still in the superconducting state at 16 T and, in the low temperature high field limit (2 K, 16 T), it changes only weakly for HT up to 695°C [$(1-F_{2K,16T})$ ~ 37-42 %] but strongly increases in the 750°C/96h wire, reaching its maximum of 50%. Despite this variation, the $T_c$ transition onsets of the different wires lie within a few tenths of a Kelvin. The 16 T $T_c$-distribution $f(T_c,16T)$ plotted in figure 17 clearly reveals the progressive increase in wire uniformity: the lower $T_c$ components of the A15 distribution are always significant and only slowly flatten as the HT continues, whereas the higher $T_c$ components gradually move to the highest $T_c$. The sudden change in the 16 T $F(T)$ curves below and above 695°C suggests that different mechanisms are playing a role in the in-field behavior. A possible explanation of this will also be presented in the discussion section.

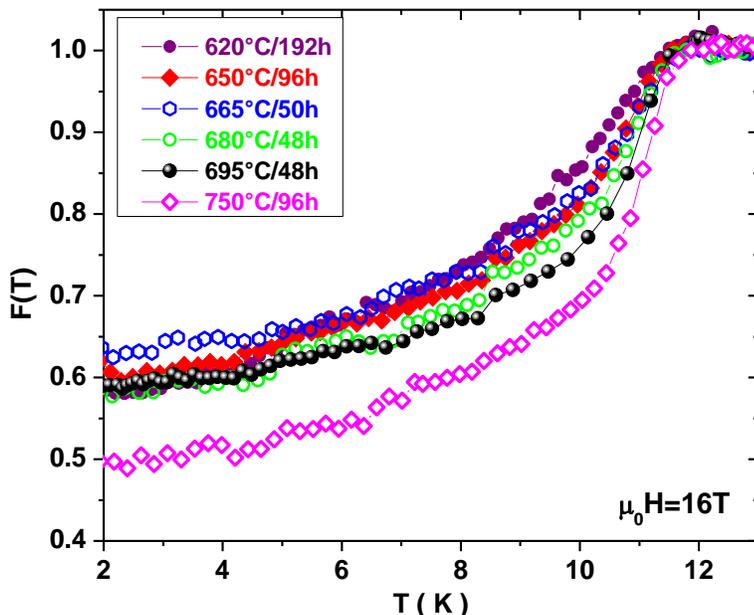

**Figure 16** $F(T)$ at 16 T showing how the transition changes in strong applied fields. The strong increase in superfluid volume fraction obtained by increasing the HT to 750°C is clear.

In summary we see that the marked signs of inhomogeneity seen in the microstructure in the last section reflect themselves quite clearly too in the $T_c$-distributions derived from specific heat. They reinforce our general conclusion that even champion $J_c$ wires like this contain a wide range of properties. The grand challenge that these results pose is of course to better minimize the distribution so as to yield yet better properties.

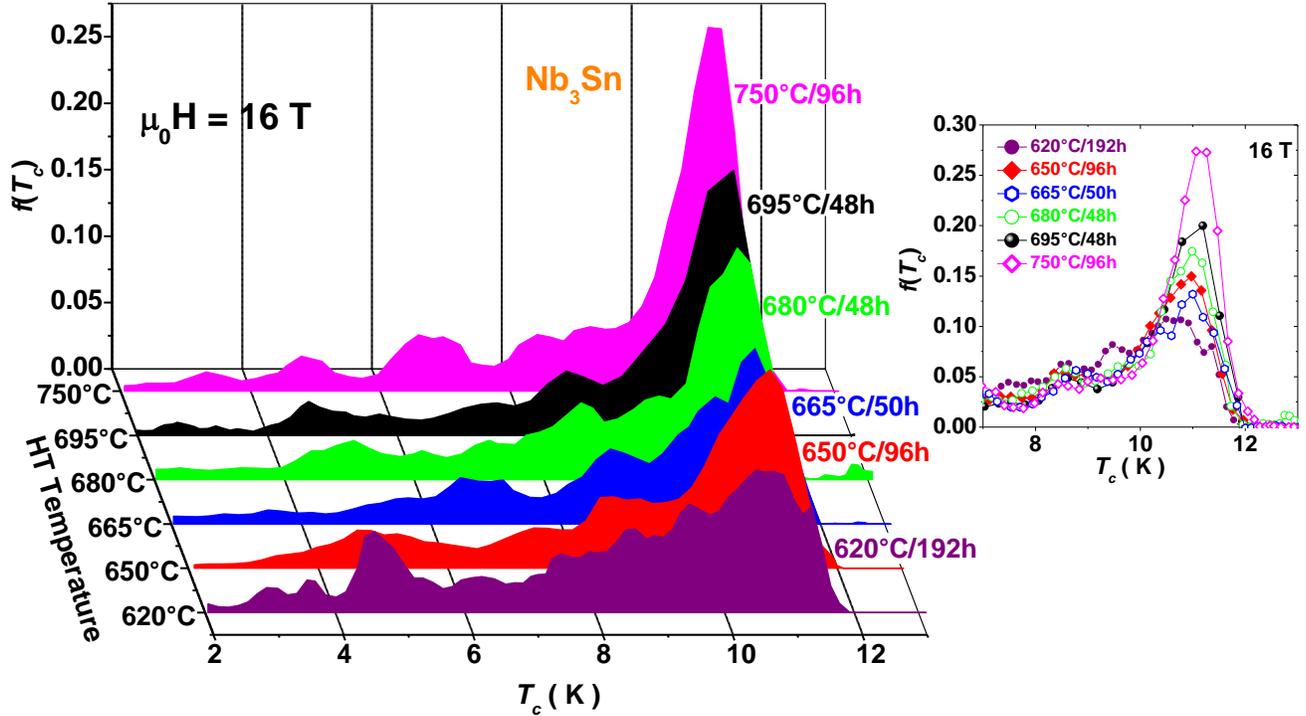

**Figure 17** $T_c$-distributions $f(T_c)$ at 16 T as a function of the heat treatment (HT) temperature and, on the right, magnified views of the highest $T_c$ A15 peak. The applied field measurements emphasize the important role of increasing HT in smoothing and pushing the $T_c$ distribution to higher temperature.

## 5. VSM characterization

### 5.1 Non-Cu cross-section properties

Magnetization hysteresis loops were measured in the VSM by applying field perpendicular to the wire axis so as to induce current flow along it. Figure 18 reports the non-Cu $J_c$ determined by the Bean model using the expression for a sub-element in perpendicular configuration [ $J_c = 15\pi * \Delta m/(4R*V)$ where $R$ is the average external sub-element radius (~35 μm), $V$ is the sub-element volume, $\Delta m$ is the total magnetic moment divided by the number of sub-elements in the wire], the pinning force density $F_p$ and the Kramer plots as a function of field at 4.2 K [30]. Since we are interested in the non-Cu $J_c$ in order to facilitate comparison with many existing transport data, the same external radius of the diffusion barrier is used in the calculation for all the wires. It is worth noting that, differently from transport measurements where normally meter-long samples are measured, in the VSM it was only possible to measure samples of a few millimeters. This means that the so-obtained non-Cu $J_c$ (and, as a consequence, $F_p$) represents the local properties of the measured pieces. For this reason and because of differences in the measurement techniques, these $J_c$ values may slightly differ from the transport results. In order to obtain consistent, although local, results we measured the same pieces of wires previously characterized by specific heat.

**Table 3** Microstructural and physical properties of the A15 wires. The percentage of A15 phase in the sub-elements (A15% 0f non-Cu) and the grain boundary density (GB density) were determined by analytical microscopy and fractography. The non-Cu values at 4.2 K were determined by VSM measurements using the Bean model (for $J_c$) and Kramer extrapolation (for $H_{Irr}$). The A15 layer $J_c$ and $F_p$ at 4.2 K were calculated rescaling the non-Cu values on the cross-section occupied by the A15 phase (A15 % of non-Cu). The GB specific pinning force, $Q_{GB}$, is determined by $Q_{GB} = F_p/\lambda S_{GB}$ (see main text).

| Heat Treatment | | A15 % of | GB density | \multicolumn{4}{c|}{Non-Cu values} | \multicolumn{4}{c|}{A15 layer values} |
| Temp °C | Time h | non-Cu | $10^7$ m$^{-1}$ | $H_{Irr}$ T | $J_c$(12T) A/mm$^2$ | $F_{p,Max}$ GN/m$^3$ | $F_p$(12T) GN/m$^3$ | $J_c$(12T) A/mm$^2$ | $F_{p,Max}$ GN/m$^3$ | $F_p$(12T) GN/m$^3$ | $Q_{GB}$(12T) N/m$^2$ |
|---|---|---|---|---|---|---|---|---|---|---|---|
| 620 | 192 | 58.8 | 2.65 | 22.71 | 3086 | 64.5 | 37.0 | 5248 | 109.7 | 62.9 | 7124 |
| 650 | 96 | 59.7 | 2.22 | 22.86 | 3067 | 63.0 | 36.8 | 5137 | 105.5 | 61.6 | 8330 |
| 665 | 50 | 59.3 | 2.11 | 22.73 | 2709 | 56.0 | 32.5 | 4568 | 94.4 | 54.8 | 7792 |
| 680 | 48 | 60.9 | 1.99 | 23.47 | 3007 | 59.2 | 36.1 | 4938 | 97.2 | 59.3 | 8936 |
| 695 | 48 | 63.7 | 1.66 | 24.81 | 2832 | 52.2 | 34.0 | 4446 | 81.9 | 53.4 | 9646 |
| 750 | 96 | 65.8 | 1.19 | 25.20 | 2097 | 38.4 | 25.2 | 3187 | 58.4 | 38.3 | 9655 |

As shown in the previous two sections, heat treatment also has a significant effect on the $J_c$ and $F_p$ behavior. The two lowest temperature HTs show only small differences and in fact they have the best performance in terms of pinning force with an $F_p$ maximum of 63-64.5 GN/m$^3$. Increasing the HT temperature to 750°C/96h, both the non-Cu $J_c$ and $F_p$ are strongly suppressed ($F_{p,Max}$ by 40%). Interestingly, $F_{p,Max}$ does not monotonically change with the HT temperature (665°C/50h has a lower $F_{p,Max}$ than 680°C/48h). This suggests that the longer HT of both 620°C/192h and 650°C/96h positively affects the $J_c$ performance more effectively than increasing the HT temperature to 665°C but for a shorter HT (only 50 h). The short HT becomes effective only with a further increase in temperature to 680°C. The extrapolated $H_{Irr}$(4.2 K), estimated by the Kramer plot, has small variations in the 620-665°C range (22.7-22.9 T) and then it significantly improves, reaching 25.2 T in the 750°C/96h wire (see table 3). Because of the long extrapolation, some inaccuracy in the $H_{Irr}$(4.2 K) values is possible, in particular if the long tail (due to inhomogeneity) frequently observed in the Kramer plot is affecting our samples [15]; however this inaccuracy is likely smaller than in many transport results where $J_c$ is usually measured up to 12 T or less. In the wires heat treated up to 695°C the position of the $F_p$ maximum is located at the same reduced field, $H_{max}/H_{Irr}$=0.19-0.2, consistent with the assumption that grain boundaries are the main pinning mechanism and that a Kramer extrapolation is valid. Only in the 750°C/96h case the ratio $H_{max}/H_{Irr}$ is slightly lower (0.18); however the Kramer plot remains linear over a wide high-field range making the extrapolation still valid. This suggests that at 750°C/96h the predominant pinning mechanism at high field is still from grain boundaries and a possible additional mechanism is effective only in the low-field region.

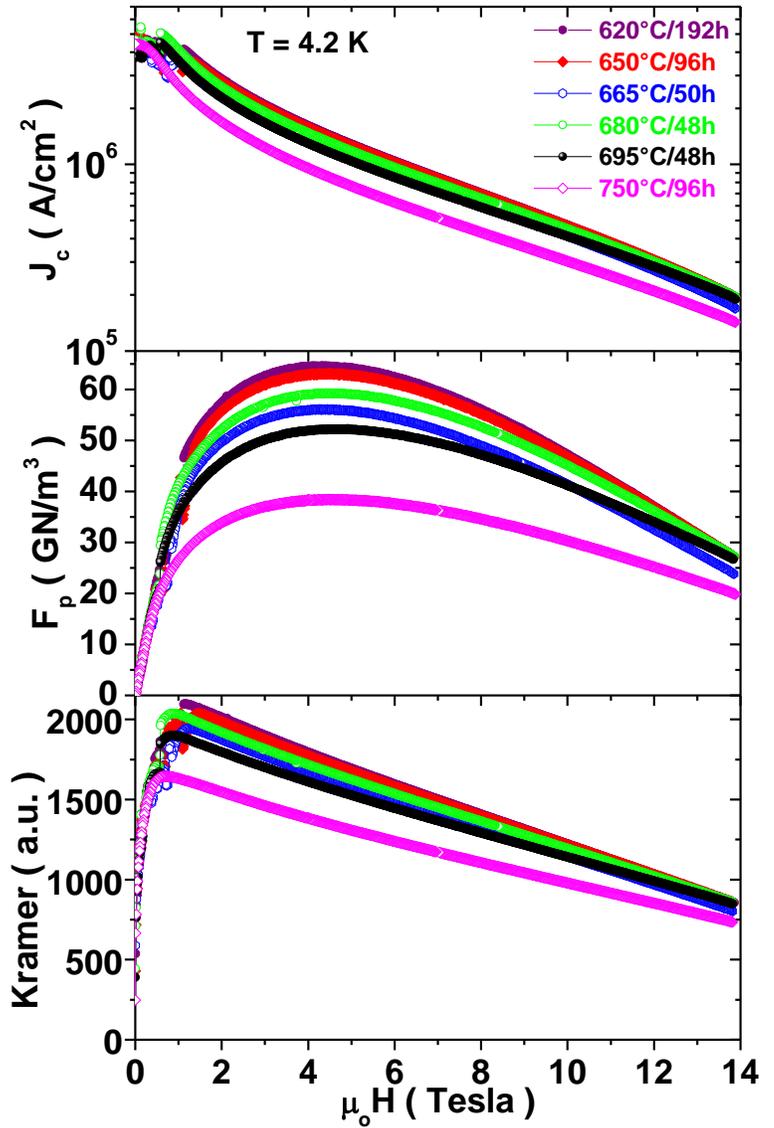

**Figure 18.** Non-Cu $J_c$, $F_p$, and Kramer plot at 4.2 K as a function of applied field for the 6 wires obtained after different heat treatment.

*5.2 A15 layer $J_c$ and pinning force properties*

Thanks to the microstructural information presented in section 3 (A15 percentage of non-Cu and GB density), the A15 layer properties can be determined as well. It is important to notice that the change in the A15 percentage with HT does not determine the large variation of non-Cu $J_c$ (40%) from HT to HT: in fact the wire with the largest non-Cu $J_c$ at 12 T (620°C/192h) also has the smallest A15 percentage. As a consequence taking into account the A15 percentage produces an even larger variation in the A15 layer $J_c$ (65%). Of particular interest is the specific GB pinning force determined by $Q_{GB} = F_p/\lambda S_{GB}$ (where $S_{GB}$ is the GB density and $\lambda$ takes into account the proportion of grain boundary oriented for pinning). The parameter $\lambda$ assumes the value 1/2 for columnar grains and 1/3 for equiaxed grains [31]. In low-Sn Nb$_3$Sn strand designs such as the bronze-process, a large portion of the A15 has columnar grains, i.e. a high aspect ratio [16,32] and $\lambda$=1/2 can be used. On the other hand, high-Sn designs such as PIT and RRP® wires have more equiaxed A15 grains (small aspect ratio) as shown in figures 8 and 9, so the use of $\lambda$=1/3 is more appropriate. We find that $Q_{GB}$ also has a strong variation with HT (see table 3): $Q_{GB}$(4.2 K, 12 T) has a minimum of ~7120 N/m² in the 620°C/192h wire, then it increases and reaches a maximum of ~9650 N/m² at 695°C/48h and 750°C/96h. Such a variation of $Q_{GB}$ suggests that the performance of the wires does not simply change because of the decreasing GB density but also because the ability of grain boundaries to pin vortices is a function of other changes produced by the different HTs as well.

## 6. Discussion

*6.1 Motivation of our study*

Our goal in this work was to try to provide a quantitative description of a real state-of-the-art, highest $J_c$, long-length Nb$_3$Sn conductor of the RRP® design so that progress towards other desirable conductor properties such as a smaller filament size, undegraded RRR and better high field $J_c$ performance could become possible and more predictable. At present RRP® and PIT conductors vie for application in the very demanding environment of the Large Hadron Collider (LHC) and the demands of the LHC are large enough to be driving much R&D on Nb$_3$Sn [1,4]. PIT and RRP® possess different architectures and reaction paths, but underlying their intensive development is a common question: is there unused space in the parameter optimization available to make them even better? This paper attempts to provide a baseline quantitative description of a large filament, high $J_c$ RRP® wire so that the reason for the lower $J_c$ and lower RRR in finer filament, greater sub-element number conductors for the same reaction can be understood. Our belief is that a full quantitative description will be able to find further optimization space to enable the desire for the very highest $J_c$ and RRR behavior to be obtained in smaller effective filament diameter versions of the state-of-the-art RRP® design.

*6.2 Summary of results and layout of discussion*

The extensive range of measurements presented here shows that even though these wires have some of the best critical current densities of any Nb$_3$Sn strands, they still have a remarkably inhomogeneous A15 layer, even when heat treated at temperatures much higher than typical. Comparison of results obtained by image analysis and specific heat with the more usual critical current characterizations provides a detailed picture of the way that intrinsic ($T_c$, $H_{c2}$, Sn content, …) and extrinsic (grain boundaries density, $J_c$, …) properties play off against each other to control the achievable properties of the reacted conductor. For this conductor at 12 T and 4.2 K, it is clear that the non-Cu $J_c$(12 T, 4.2 K) properties are still strongly coupled to the A15 grain size and that higher temperature reactions degrade both the non-Cu $J_c$(12 T, 4.2 K) and the RRR. For present accelerator use, where peak fields are in the 12 T range, lower temperature reactions (which emphasize a high A15 grain boundary density) rather than higher temperature reactions (which push the A15 layer composition more towards the stoichiometric composition) win out, showing that extrinsic rather than intrinsic property optimization is paramount. But many uses of Nb$_3$Sn demand optimization at fields of 16 T or more. Most Nb$_3$Sn laboratory magnets have peak fields in the 16-20 T range, thus focusing attention on the value of raising the in-field superfluid volume fraction by more successfully homogenizing the chemistry, in this case by increasing the reaction temperature.

The detailed microstructural and chemical analyses, the zero-field and in-field 16 T $T_c$-distributions and the $J_c$, $F_p$ and $Q_{GB}$ tabulations reveal many interesting and complementary correlations. In this section we now discuss the relevance of our findings for the diffusion barrier (section 6.3), the comparison of specific heat and magnetization measurements (section 6.4), the value of increasing the reaction temperature and its inability so far to avoid local gradients within each filament, as well as across the filament pack within each sub-element (section 6.5), the correlation of the in-field $T_c$-distributions and superfluid fraction extracted from specific heat to the irreversibility field (section 6.6) and the critical current density (section 6.7). We close (section 6.8) with some comparison to other conductor types and suggestions for small changes to the design of RRP® conductors.

*6.3 Diffusion barrier properties and maintenance of high RRR in the stabilizing Cu*

The quantitative microscopy reveals a progressive decrease in the amount of unreacted Nb-Ta barrier from 8.1 to 3.7% of the non-Cu cross-section (table 1). A positive observation is that RRR > 100 occurs when only 5.6% of residual diffusion barrier remains, allowing almost 61% of the non-Cu cross-section to convert to the A15 phase. This is a much more efficient design than occurs in PIT conductors where it appears that about 25% of residual barrier is needed to maintain the same RRR [33]. Figure 4 shows that some react-through of the A15 to the Cu occurs at every temperature but below 665°C it is limited to <0.2%, whereas it reaches 23% of the total barrier perimeter at 750°C. As a consequence, sufficient Sn diffuses into the Cu stabilizer to form A15 layers outside the barrier at higher temperatures (e.g. the 750°C/96h wire in figure 2) compromising the RRR of the stabilizing Cu, as revealed by the degradation of the RRR to unacceptable values below 100 for heat treatment above 695°C (figure 4 and table 1). A key finding of the full digital analysis shown in figure 4a is that RRR falls to ~100 when only ~2% of the barrier is reacted through. Figure 4 also shows that this barrier breakdown occurs preferentially in outer ring sub-elements, particularly in the outer-ring corner elements where sub-element distortion is greatest (figures 1 and 6). These results strongly suggest that better conductor architectures, especially those with a greater number of smaller sub-elements that are now being favored, must control this local breakdown. Degraded RRR is not at all a global property, but one triggered by very localized breakdown. Comparison to a study of a recent 192 filament PIT design with about 25% residual barrier (admittedly with 50 μm diameter rather than the 70 μm diameter sub-elements studied here) suggests both that RRP® conductors could be optimized by somewhat thicker barriers, while PIT conductors might develop higher non-Cu $J_c$ with somewhat thinner barriers.

For single strand magnets, for example most solenoids, this finding emphasizes that an important route to greater protection of the stabilizing Cu would be by better control of the deformation of the outer ring sub-elements and perhaps by preferential thickening of the barriers in those sub-elements that form the outer corners of the outer ring. We also have recently seen that similar greater deformation of outer ring filaments occurs in round, as-drawn PIT conductors too [33]. By contrast studies of rolled round wires aimed to simulate Rutherford cabling damage tend to show that the worst damage occurs for inner ring filaments as macroscopic shear bands distort the filament pack [34, 35, 36]. In any case the present results make it clear that RRR loss is caused by very local barrier breakdown, making diffusion barrier thickness a vital control parameter. For the specific case of 12 T use for accelerators, actually a reaction of up to 48 h at 680°C satisfies both RRR and $J_c$(12 T, 4.2 K) > 3000 A/mm$^2$. Only for higher field use does the barrier breakdown of this conductor prevent more beneficial, higher temperature reactions. Smaller sub-elements just reinforce this problem. In short it seems that diffusion barrier breakdown is always a key, real conductor problem. A distinctive finding of this study is that we have provided a level of detail about the diffusion barrier breakdown that is quite new.

*6.4 Comparison of the specific heat and magnetization measurements*

In contrast to the metallographic measurements, the specific heat analysis does not provide location information but it does provide a true bulk average of the $T_c$ variation on a scale of the coherence length and clearly confirms the consumption of the Nb-Ta diffusion barrier by a progressive decrease in the $T_c$-distribution of the amplitude of the Nb peak (figure 15), which almost disappears at 750°C.[a] Its real value is however to show the great inhomogeneity of the A15 $T_c$-distribution with a finer resolution and without any screening artifacts as can and generally does occur in magnetization measurements. The $T_c$-distributions $f(T_c)$ shown in figure 15 make it clear that, even in this example of the highest $J_c$ conductor yet made, major residual compositional and superconducting inhomogeneities remain.

---

[a] The difference in the estimation of Nb amount obtained by the specific heat and microstructural analyses is due to the fact that the model of ref. 16 assumes one unique Sommerfeld constant γ in eq.1 despite the presence of Nb and the A15 and a gradient of the superconducting properties in the A15 phase. Moreover the specific heat analysis cannot distinguish between the Nb in the diffusion barrier and in unreacted filament cores, as shown in figure 6 and 7. More details will be included in a paper in preparation.

Concerning screening and other artifacts, the magnetization measurements in figure 14 show a lower $T_c$-onset than the $F(T)$ curves, which we interpret as being due to the disconnected nature of the highest $T_c$ regions which cannot produce a significant magnetization screening signal, perhaps due to the shape irregularities of the inner A15 ring that forms at the Sn core-filament pack interface. There may also be size effects too, since magnetization detects on a scale of a few penetration depths $\lambda$ ($\lambda(0) \sim 200$ nm), while the specific heat detects on the scale of $\xi$ ($\xi(0\ K) \sim 3$ nm). Screening effects of the magnetization measurement are indicated by the similarity of the relative amplitudes of the Nb and A15 transitions after different reaction (inset, figure 14), whereas the integral of the $T_c$-distribution, $F(T)$, much more clearly reveals the progressive reaction of the Nb barrier to A15 phase, some of which has $T_c$ lower than that of the unreacted barrier. Shielding by the outer section of the Nb-Ta barrier hides this low-$T_c$ A15 in magnetization measurements, leaving the region screened by Nb substantially independent of the extent of the A15 reaction. Moreover the $M(T)$ curves in figure 14 reveal different $T_c$ values for the two samples shown but similar widths, whereas the $F(T)$ curves clearly show a strongly different broadening of the A15 transitions, indicating again that some of the low-$T_c$ A15 is screened by higher-$T_c$ A15 in the magnetization measurement (probably at the level of the original Nb(Ta) filaments, see section 6.5). In summary it is clear that the (much quicker) magnetization measurements can yield valuable information, but also that specific heat measurements are more accurate. The two together can be particularly powerful since the completely location-independent averaging of the specific heat can be compared to the partial location-dependent magnetization measurements.

*6.5 Compositional and superconducting inhomogeneity*

Greater A15 compositional homogeneity can be achieved either by longer heat treatment at lower temperatures or by shorter higher temperature heat treatments. As a practical matter, however, both specific heat and EDS measurements of local composition show clearly that no reaction used here can completely homogenize the A15 phase of this conductor. Figure 10 shows the Sn composition across the A15 layer with changing HT. This chemical analysis shows the compositional inhomogeneity in the A15 phases in the 620°C/192h wire (particular evident in the external portion of the A15 layer, far from the Sn source). In the 680°C/48h sample, a Sn gradient is still present but with a much smaller filament-to-filament variation. Despite reduction of the Sn gradient with increasing HT temperature, the FESEM-EDS mapping reveals spatial compositional variations at the filament level (figure 11) with low-Ta and high-Sn rings present even after 750°C reaction. RRP® wires have two gradient components, one being the radial Sn gradient outwards across the sub-element from the core [37, 38], the second being a more local radial gradient within each filament (this also explains why the A15 $F(T)$ transitions are significantly broader than the $M(T)$ transitions). Moreover Ta segregation is drastically reduced on increasing reaction from 695°C to 750°C. All these inhomogeneities, multiple composition gradients and their variation with location and heat treatment are what generate the strong $T_c$-distributions found by specific heat. Since $T_c$ of the A15 phase is related to the Sn content [7], the change in width and amplitude of the $F(T)$ transition (figure 14) reflects a flattening Sn compositional gradient as the HT is extended. For instance in the 620°C/192h wire the large compositional inhomogeneity (figure 11) and strong filament to-filament variation (figure 10) produces a smeared and the smallest peak in the $T_c$-distribution (figure 15). The specific heat analysis detects also small differences that are not easily determined by FESEM imaging or EDS analysis. In fact the $F(T)$ and $f(T_c)$ comparisons between 650°C/96h and 665°C/50h suggest that the longer, lower temperature HT (650°C/96h) is more effectively driving Sn into the whole annulus of each sub-element, thus producing a more homogeneous A15 layer with narrower transitions ($\Delta T_c \sim 3$ K versus 3.7 K for 665°C/50h). By contrast, the only indication of this conclusion in the image analysis is just a slightly smaller A15 percentage in the sub-element of 665°C/50h with respect to 650°C/96h. For the short heat treatments (48-50 h) only a further increase of the HT temperature up to 680-695°C produces enough Sn diffusion to further narrow the $F(T)$ transition from 3.7 K at 665°C to 2.3 K at 695°C.

## 6.6 In-field $T_c$-distributions, superfluid fraction and irreversibility field

The $F(T)$ curves obtained at 16 T (figure 16) highlight even more small changes to the overall A15 inhomogeneity. In fact the transitions clearly sharpen with increasing HT temperature, as more Sn is driven into the filament pack, the only exception being the shorter 665°C/50h that again results in lower homogeneity than the longer 650°C/96h HT. This is also evident in the smaller $f(T_c,16T)$ peak (figure 17) for 665°C/50h compared to 650°C/96h. Another somewhat unexpected result in the $F(T,16T)$ curves is that raising the heat treatment to 695°C has little effect on the 2 K and 16 T superfluid volume fraction (1-$F_{2K,16T}$), while raising it to 750°C provides a sudden increase with 1-$F_{2K,16T}$ reaching to 50%. Changes in the spatial compositional segregation at the filament scale may provide an explanation. As mentioned before, besides a further decrease in the Sn gradient, a notable homogenization of the Ta within the filament is also observed at 750°C. Since the $H_{c2}$ of the A15 phase is strongly dependent on the Ta doping [39], a more uniform Ta distribution in the high-Sn ring of A15 around the outer radius of each former filament may positively affect the in-field $T_c$-distribution.

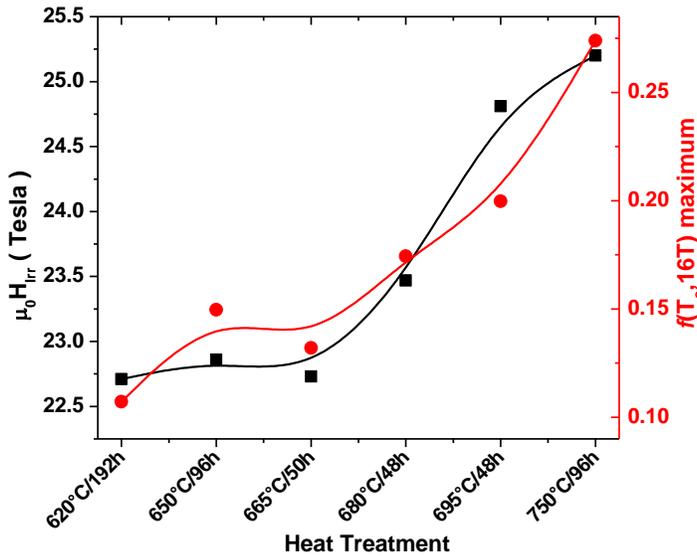

**Figure 19** Extrapolated irreversibility field $H_{Irr}$ at 4.2 K and the maximum of $f(T_c,16 T)$ as a function of HT. $H_{Irr}$ and $f(T_c,16 T)$ show a similar trend.

Correlations are also found between the in-field $T_c$-distribution and the irreversibility field determined by the Kramer extrapolation. In fact, as shown in figure 19, 620°C/192h and 665°C/50h have the least intense peaks, followed by 650°C/96h and the higher temperature heat treated wires. This same trend is followed by the extrapolated $H_{Irr}(4.2 K)$, also reported in figure 19. Because of the weak thermal fluctuations in the LTS, $H_{c2}$ and $H_{Irr}$ are closely related: an enhancement of $H_{c2}$ produces a proportional enhancement of $H_{Irr}$. Since $H_{c2}$ rapidly increases with increasing $T_c$ [12,40], the narrowing and shift of the $T_c$-distribution (both at 0 and 16 T) to higher temperature implies a closely correlated shift of $H_{c2}$ and $H_{Irr}$ distributions. This was already noted in an earlier detailed study of the $H_{Irr}$ and $H_{c2}$ distribution in PIT conductors with varying HT conditions [12].

## 6.7 In-field $T_c$-distributions, superfluid fraction and critical current density

In contrast to the 12 T optimizations discussed earlier, where the most inhomogeneous HT made at 620 °C produced the highest $J_c$, our results also show that enhanced A15 homogeneity obtained by higher HT temperature can improve the very high field $J_c$ performance because $H_{Irr}$ and in-field $T_c$-distribution correlate. Note that the non-Cu $J_c(H)$ of 695°C/48h crosses the other curves above 10 T (figure 18), although it is not very effective at intermediate field where $J_c$ (and $F_p$) is most directly controlled by the A15 grain size. In fact the largest $J_c(12 T)$ and $F_p$ maximum is found in the wire heat treated at the lowest temperature (620°C) where the GB density (and the pinning centers that they provide) is at a maximum. The delicate interplay between A15 grain boundary density and A15 compositional homogeneity is particularly evident in comparing the 650°C/96h, 665°C/50h and 680°C/48h samples. In fact the grain boundary density slightly decreases (table 3) with increasing HT temperature but it is the 665°C/50h sample that has the lowest $J_c$ (and $F_{p,Max}$). This variation demonstrates that $J_c$ (and $F_{p,Max}$) is indeed determined both by optimization of the $T_c$-distribution (both at 0 and 16 T) and the A15 phase homogeneity, not just by maximizing the GB density. The higher the field at which $J_c$ should be optimized, the more compositional and superfluid volume fraction optimization wins out over prevention of A15 grain growth. Another potentially important factor is that the grain boundary pinning strength, $Q_{GB}$, strongly increases with increasing HT temperature [$Q_{GB}(4.2 K, 12 T)$ rises by about 35% on going from 620 to 750 °C]. Whether this is because of the variations in grain boundary chemistry is unclear, but it is clear that $Q_{GB}$ is a function of HT temperature too.

*6.8 Comparison to other conductor designs and some broader implications*

In terms of our qualitative conclusions, this study does not break new ground, since all of the main themes of the effect of HT temperature and extent on composition of the A15 and the $J_c(H)$ have been found before [6-8,11-12,15-17]. PIT [18, 19, 41, 42, 43] and Sn-in-tube composites [44, 45, 46] have been particularly valued for study because there is only an outward radial Sn flux from each Sn-rich core into the surrounding tubular sub-element. For instance, Hawes *et al.* [18,19] and Godeke *et al.* [12] made extensive study of a wide range of HT in PIT conductors, coming to qualitatively similar conclusions. But we have recognized for some time that we needed a quantitative outcome. An earlier attempt was with the quantitative compositional shell model developed by Cooley *et al.* [15], which enabled the effect of variable non-linear Sn gradients on $J_c$ and $H_{Irr}$ to be predicted. More recently Senatore and Flükiger [27] investigated a PIT conductor with two distinct specific heat $T_c$ distributions which they attributed to fine and coarse grain A15 regions. Inhomogeneity is thus endemic to all modern designs of A15 conductor. What is in general missing from these earlier studies are specific recommendations for how to drive conductors towards the target that modern accelerator projects really want, namely $J_c(12T, 4.2K) > 3000$ A/mm$^2$, $J_c(15T, 4.2K) > 1500$ A/mm$^2$ at $d_{eff} < 20$ μm [1].

In the RRP® wires, we assert that it is reasonable to expect more complexity to the compositional (and $T_c$) variations than in the case of a single-filament design like PIT. What this study emphasizes is that an important obstacle to finer sub-elements is localized diffusion barrier breakdown. Smaller sub-element diameters will certainly require relatively thicker barriers while higher field use will increase the importance of developing a higher in-field superfluid volume fraction, even if the grain boundary density falls. High field solenoid magnets, an important market today, generally have sections operating at fields in the 16-20 T range, where optimization for maximum $H_{irr}$ by higher temperature reaction would seem called for. Recent developments of RRP® conductors [2] have tended to show that smaller sub-elements lead both to degraded RRR *and* lower $J_c$. Quantitative analysis of the sort presented here should be able to sort out whether barrier breakdown is a local problem as found here and how $J_c$ is determined by the balance between intrinsic and extrinsic factors. Although this requires an extensive characterization, it is our conviction that this kind of quantitative evaluation is valuable for comparing small differences in design of the sort that are now characteristic of Nb$_3$Sn conductor development. This kind of investigation may suggest how to decrease the double composition gradient observed here. Decreasing both gradients, while limiting A15 grain growth might be enabled by two-step heat treatments similar to those now being performed on PIT wires [47] and commonly used in bronze route wire [48,49]. Understanding the compromises in all present state-of-the-art Nb$_3$Sn wires is clearly not a qualitative issue but a highly detailed quantitative issue, where choices have to be made between optimizing the intrinsic properties such as $T_c$ and $H_{c2}$ at the expense of extrinsic properties such as GB density and diffusion barrier integrity. Our goal with this study was to provide a more detailed and overlapping set of characterizations capable of taking Nb$_3$Sn to a new level of performance.

**7. Conclusions**

Here we have coupled specific heat measurements with detailed microstructural, microchemical and electromagnetic characterizations of a state-of-the-art, very high $J_c$ 54/61 RRP® Nb$_3$Sn conductor so as to better understand where the freedom in the processing space is. Our goal in making this very broad and quantitative characterization was to find ways to allow even higher $J_c$, smaller sub-elements without loss of RRR. We found that increased time and temperature of the heat treatment markedly enhanced the compositional homogeneity of the A15 phase without however removing major inhomogeneities, which remain even after HT at 750°C that completely destroys the diffusion barrier. Such higher temperature reactions raise the in-field superfluid fraction, the irreversibility field and the highest field $J_c$. At 12 T, 4.2 K, however, minimizing A15 grain growth by restricting the HT is as well-known still important. For this 70 μm diameter sub-element design, diffusion barriers can prevent RRR falling below 100 up to reactions of 48h at 680 °C. Detailed analysis of the barrier shows that RRR ~ 100 corresponds to only ~2% of the barrier being breached by the A15 reaction front. This emphasizes how essential it is to concentrate diffusion barrier studies on all incipient breaks, rather than concentrating on average properties.

As work continues to optimize conductors of this RRP® design, it is clear that stronger barriers will be vital in driving from the 61 sub-element design used here to the smaller sub-element 127, 169 and 217 stack designs. The target field range will also play a large role in selecting a reaction temperature and barrier quality. Higher field use is favored by longer and hotter reactions that place more stress on the diffusion barrier. The opportunity for small changes of design or fabrication sequence to enhance properties is suggested by the complex variations of chemical composition seen in the A15 layer of this very highest $J_c$ conductor. A macroscopic gradient from Sn core to sub-element diffusion barrier is seen for all HT types. Some locally deformed regions appear to resist local reaction of filaments, suggesting proximity to a pinch off of the Sn diffusion flux necessary to convert the Nb-Ta filaments to A15. A second quite general gradient was seen on the scale of each original filament, even though almost all of the original Cu surrounding each filament in fabrications counter-diffuses with the Sn, ending up in the core.

We intended this characterization to be broad and to serve as a benchmark for further characterizations of finer sub-element RRP® conductors. Since the high non-Cu $J_c$(12 T, 4.2 K) values measured here are obtained with about 60% A15 and about 5% of residual diffusion barrier, there does appear to be considerable room for development of stronger barriers capable of allowing finer sub-elements with high $J_c$. Since high $J_c$ Nb$_3$Sn conductor design is still evolving, the analysis method presented here may be of use in evaluating such conductors quite generally. Our study suggests that all such high $J_c$ conductors are markedly inhomogeneous and that optimization must pay proper attention to the nature and location of these inhomogeneities.

**Acknowledgments**

This work was supported by the US Department of Energy (DOE) Office of High Energy Physics under grant number DE-FG02-07ER41451, by the National High Magnetic Field Laboratory (which is supported by the National Science Foundation under NSF/DMR-1157490), and by the State of Florida. Work at BNL is supported by the US Department of Energy under Contract No. DE-AC02-98CH10886. The wire was developed under the DOE Conductor Development Program managed by D. Dietderich of the Lawrence Berkeley National Laboratory. Many discussions with Michael Brown, Chris Segal and Zu-Hawn Sung are gratefully appreciated.

**Appendix**

In order to investigate the superconducting properties, the electronic contribution to the total specific heat of the superconducting phases alone has to be isolated, taking into account the presence in the wire of the two superconducting materials, Nb$_3$Sn and Nb, and normal materials, like Cu and other Nb-Sn-Cu phases (for simplicity in the following equations the contributions from the superconducting materials will be indexed as "Nb$_3$Sn" and those from the normal ones by "Cu"). Following the usual procedure described in ref [29,16,17], the normal-state specific heat is determined by fitting the 16 T data above 12-12.5 K (the onset of the A15 transition at 16 T) with the relation $C_{fit}(T) = \gamma_{fit}T + \beta T^3 + \delta T^5$ and verifying the conservation of the entropy above $T_c$ [28, 16]: $\int_0^T \frac{C_{fit}(t)}{t} dt = \int_0^T \frac{C(t, B=0)}{t} dt$ for T $\geq$ $T_c$. As well known, the phonon contribution $C_{ph}(T) = \beta T^3 + \delta T^5$ (that in this case is also equal to $C_{ph,Nb_3Sn}(T) + C_{ph,Cu}(T)$) is not affected by the occurring superconducting transition. As regards the electronic contributions, the one for the normal phases is determined by $\gamma_{Cu}T$ at every temperature, whereas the one from the superconducting phases is given by $\gamma_{Nb_3Sn}T$ above the $T_c$ onset and it assumes a more complicated temperature dependence below $T_c$ (as described in the text). As a consequence, above the $T_c$ onset the total electronic contribution is given by $\gamma_{Nb_3Sn}T + \gamma_{Cu}T = \gamma_{fit}T$ (with $\gamma_{fit}$ from the normal-state fit above). $\gamma_{Cu}$ can be estimated by the zero-temperature extrapolation of the low temperature data in the $C/T$ versus $T^2$ plot, since $(C/T)_{Nb_3Sn}$ tends to zero whereas $(C/T)_{Cu}$ tends to $\gamma_{Cu}$. Finally, the electronic contribution of the superconducting phases only, $C_e$, is determined by subtracting both the phonon contribution and the electronic contribution on the normal phases from the total specific heat: $C_e(T) = C(T, B=0) - C_{ph}(T) - \gamma_{Cu}T$. The further analysis described in the main text following Wang's paper [16] will use the Sommerfeld constant of the superconducting phase only ( $\gamma = \gamma_{Nb_3Sn}$ ).